\documentclass[a4paper,11pt]{article}
\pdfoutput=1 

\usepackage{jheppub} 

\usepackage[T1]{fontenc} 

\usepackage{amsmath,amssymb,amsfonts}
\usepackage{graphicx}
\usepackage{hyperref}
\usepackage{bbm}
\usepackage{mathdots}
\usepackage{color}
\usepackage{ytableau}
\ytableausetup{aligntableaux=center,smalltableaux}
\usepackage{tabularx}
\usepackage{here}
\usepackage{comment}

\usepackage[nameinlink]{cleveref}
\crefname{section}{}{}
\creflabelformat{section}{#2Sec.~#1#3}
\crefname{equation}{}{}
\creflabelformat{equation}{#2Eq.~(#1#3)}

\newcommand{\Yfund}{\ydiagram{1}}
\newcommand{\Yasymm}{\ydiagram{1,1}}
\newcommand{\Ythreea}{\ydiagram{1,1,1}}

\def\cW{\mathcal{W}}

\def\bZ{\mathbb{Z}}

\def\tr{\mathrm{tr}}
\def\adj{\mathbf{adj}}
\def\det{\mathrm{det}}
\def\Pf{\mathrm{Pf}}
\def\one{\mathbf{1}}

\usepackage{cellspace}
\newcommand{\myspace}[1]{
    \setlength{\cellspacetoplimit}{#1mm}
    \setlength{\cellspacebottomlimit}{#1mm}
}

\title{\boldmath Truly Confining Supersymmetric Gauge Theories}

\author[a]{Riku Ishikawa,}
\author[a,b,c,1]{Hitoshi Murayama,\note{Hamamatsu Professor}}
\author[a]{Shota Saito}

\affiliation[a]{Kavli Institute for the Physics and Mathematics of the Universe (WPI), UTokyo Institutes for Advanced Study, University of Tokyo, Kashiwa 277-8583, Japan}
\affiliation[b]{Department of Physics, University of California, Berkeley, California 94720, USA}
\affiliation[c]{Theoretical Physics Group, Lawrence Berkeley National Laboratory, Berkeley, California 94720, USA}

\emailAdd{riku.ishikawa@ipmu.jp}
\emailAdd{hitoshi@berkeley.edu}
\emailAdd{shota.saito@ipmu.jp}

\abstract{We classify ``truly confining'' ($t$-confining) supersymmetric gauge theories, in which no center charges can be screened, and Wilson loops in the fundamental representation are therefore expected to exhibit an area law. In all cases, we identify the condensation of certain ``magnetic'' operators. Many of them have more than three branches, and one with vanishing superpotential, a phenomenon not previously observed.}

\begin{document} 
\maketitle
\flushbottom

\section{Introduction}
\label{sec:intro}

Quark model was originally proposed by Gell-Mann \cite{Gell-Mann:1961omu} and Zweig \cite{Zweig:1964ruk,Zweig:1964jf} to categorize a zoo of hadrons observed experimentally in representations of flavor $SU(3)$ as bound states of up, down, and strange quarks. An obvious problem was that quarks were never observed, and they were considered as mathematically useful construct but not real particles. If taken seriously, a big assumption had to be made that quarks were {\it confined}\/ by some dynamics. This assumption was grudgingly accepted because it was required by later experimental observations. Deep Inelastic Scattering (DIS) experiments showed that there are nearly free ``partons'' inside proton \cite{Bloom:1969kc,Breidenbach:1969kd}. Asymptotic freedom in non-abelian gauge theories with a negative one-loop beta function \cite{Gross:1973id,Politzer:1973fx} provided a possible explanation for the nearly free behavior at high energies. The dramatic discovery of $J$ at Brookhaven \cite{E598:1974sol} and $\psi$ at SLAC \cite{SLAC-SP-017:1974ind} as an extremely narrow resonance supported the idea that it is a $c\bar{c}$ bound state that is narrow because of the small coupling $\Gamma \propto \alpha_s^3$ \cite{Appelquist:1975ya}. Finally observation of ``naked charm'' mesons such as $D^0 = (c\bar{u})$ \cite{Goldhaber:1976xn}, $D^+ = (c\bar{d})$ \cite{Peruzzi:1976sv} convinced the community that the charm quark exists, together with up, down, and strange as ``real'' constituents of hadrons. Yet the idea of confinement appeared mysterious.

In restrospect, there were some facts that supported the idea. The Chew--Frautschi plot \cite{Chew:1962eu} showed that the leading Regge trajectories for mesons and baryons fall on straight lines on the $(J, m^2)$ plane. It was later proposed by Nambu as a consequence of a linear potential, or equivalently a relativistic open string with energy $E = \alpha' r$ with a quark and an anti-quark attached on either end. In $1+1$ dimension, the Coulomb potential in $U(1)$ gauge theory is actually linear $V=\frac{e^2}{2}|r|$, and a heavy electron $m_e \gg e$ is confined. Schwinger showed that even massless electron cannot be isolated \cite{Schwinger:1962tn,Schwinger:1962tp}. Wilson used Euclidean lattice gauge theory in the strong coupling expansion in $3+1$ dimension that the Wilson loop can have the area law $\langle {\rm Tr} e^{i \oint A}\rangle \sim e^{-(\alpha' r) T }$ \cite{Wilson:1974sk} where $V=\alpha' r$ can be interpreted as the linear potential. The exponent is given by the product of $r$ and $T$, which demonstrates the area law. It is still regarded today that the area law of the Wilson loop is the clear order parameter of the confinement. He noted, however, the result was not Lorentz-invariant and left the possibility that the area law was an artifact of non-Lorentz-invariant latticization of spacetime. 

Clearly more robust demonstration of confinement in non-abelian gauge theories was needed. What made the discussion trickier was the realization that a massless QCD with $SU(3)$ gauge group and quarks in the fundamental representation cannot exhibit the area law of the Wilson loop. This is because the linear potential can be cut by the Schwinger process of creating a quark anti-quark pair in the strong ``electric'' field. The color charges are rather ``screened'' than ``confined'' which is the case with massless Schwinger model of $1+1$D $U(1)$ gauge theory \cite{Iso:1988zi}. 

It was realized that the area law of Wilson loops is possible only when dynamical matter fields in the theory cannot screen all center elements of the gauge group. For $SU(N_c)$ gauge groups, the center group is ${\mathbb Z}_{N_c}$. Quarks in the fundamental representation can screen any color charges because it is a faithful representation of the center group. On the other hand in the absense of matter fields, the gauge fields themselves are in the adjoint representation which is not a faithful representation of the center group. It represents $PSU(N_c) = SU(N_c) / {\mathbb Z}_{N_c}$. Then the color charge in the fundamental representation cannot be screened and we can expect the area law of the Wilson loops in representations that transform non-trivially under the center group. This is now better understood in terms of line operators \cite{Aharony:2013hda} and their one-form symmetries \cite{Gaiotto:2014kfa}.

We are therefore led to consider theories that are ``truly confining.'' Namely the non-abelian gauge theories where the matter fields transform trivially under the center group or its non-trivial subgroup. On the other hand, we still lack analytical tools to study non-perturbative dynamics of non-abelian gauge theories other than numerical simulations on Euclidean lattice. 

This is why we study ``truly confining'' ($t$-confining) supersymmetric gauge theories in this paper. Supersymmetry allows us to solve the non-perturbative dynamics of gauge theories {\it exactly}\/ in the sense that we can work out the infrared (IR) behavior of the theories about the symmetry of the ground state and low-lying spectrum of excitations. We present a classification of all supersymmetric gauge theories that truly confine. We were particularly inspired by the remark by Intriligator and Seiberg \cite{Intriligator:1994sm} about the gaugino condensation $\langle \lambda \lambda \rangle = \Lambda^3 e^{2\pi i k / h} \neq 0$ in pure SUSY Yang--Mills theories where $h$ is the dual Coxeter number of the gauge group. It is an order parameter of the breaking of the discrete ${\mathbb Z}_{2h}$ $R$-symmetry to is ${\mathbb Z}_2$ subgroup, and can be considered as an order parameter of confinement. The operator can be regarded as a ``magnetic'' object that causes the confinement. In fact, we identify condensation of ``magnetic'' objects as a common feature among all of them. 

The $t$-confining theories should not be confused with $s$-confining theories that are described by an IR-free theory among gauge-invariant polynomials. They were classified by Cs\`aki, Schmaltz, and Skiba \cite{Csaki:1996sm,Csaki:1996zb}. 
We went through their list of $s$-confining theories, and found none of them leave a center symmetry unbroken. We will show this point in Appendix. \ref{sec:sconfining}.

In this paper, we show that the complete list of $t$-confining supersymmetric gauge theories is the following.
\renewcommand{\labelenumi}{(\Alph{enumi})}
\begin{enumerate}
\item Pure Yang--Mills theories
\item $SO(k)$ gauge theories with $k-4$ flavors in the vector representation
\item $SO(k)$ gauge theories with $k-3$ flavors in the vector representation
\item $SU(6)$ gauge theory with one field in the rank-three anti-symmetric tensor representation
\item $Sp(k)$ gauge theories with one field in the rank-two anti-symmetric tensor representation
\item $SU(2k)$ gauge theories with one field in the rank-two anti-symmetric tensor representation and another in its conjugate representation
\item $SO(12)$ gauge theories with two flavors in the spinor representation
\end{enumerate}
(D) is the only chiral gauge theory on this list. The theories (E) were studied only partially for $k=2$ when equivalent to $Sp(2) \simeq SO(5)$ \cite{Intriligator:1995id}, and the theories (F) and (G) have not been studied in the literature before. Even for known theories such as (A) \cite{Veneziano:1982ah,Witten:1982df}, (B), and (C) \cite{Intriligator:1995id}, we show their properties under the discrete symmetries that demonstrate the condensation of ``magnetic'' objects. It is also interesting that many of them exhibit multiple branches. So far only theories with two (B,C) \cite{Intriligator:1995id} or three (D) \cite{Csaki:1997cu} branches of the moduli space are known in the literature to the best of our knowledge, while theories (E) and (F) can have many more inequivalent branches.


The organization of this paper is as follows.
In Sec. \ref{sec:definition}, we consider the conditions for $t$-confinement.
We essentially define $t$-confinement through the unscreening of Wilson lines, but we also impose additional conditions.
When discussing confinement, it is believed that the necessary condition is that the Dynkin index of the matter fields are sufficiently small and that the anomalies match.
In Sec. \ref{sec:classification}, we classify $t$-confining theories for simple gauge groups with no tree-level superpotential.
It is found that all $t$-confining theories can be classified into six series.
Four of these have been analyzed previously, while two are models that we analyze in this work.
In Secs.~\ref{sec:pureYM}, \ref{sec:SO(N)+(N-4)V}, \ref{sec:SO(N)+(N-3)V}, and \ref{sec:SU(6)+A}, we briefly revisit the four previously analyzed theories from the perspective of $t$-confinement.
In Secs.~\ref{sec:Sp(2k)+A} and \ref{sec:SU(2k)+A+A}, we analyze the new models (although some special cases have been considered in \cite{Cho:1996bi,Csaki:1996eu}).
In these sections, we derive the superpotentials that are dynamically generated and verify their consistency.
The features of each model as $t$-confining theories are also discussed.
In Sec. \ref{sec:SO(12)+2s}, we analyze another model of the $t$-confining theory.
We check the anomaly matching conditions and show the existence of the condensating operator.
In Sec. \ref{sec:discussions}, we reflect on the entire family of $t$-confining theories.
As the analysis progresses, the characteristics and common features of $t$-confining theories emerge, and these are summarized again.
Appendix \ref{sec:calculation} contains the calculations necessary for Secs.~\ref{sec:Sp(2k)+A} and \ref{sec:SU(2k)+A+A}.
Note that it does not cover all calculations.
Appendix \ref{sec:data_of_Lie_algebra} summarizes the Lie algebra information needed in the main text.

\section{Conditions for $t$-confining Theories}
\label{sec:definition}

In this section, we define, classify, and analyze supersymmetric gauge theories that have a truly confining phase.
In simple terms, these are gauge theories with a non-trivial center symmetry, where Wilson lines in the fundamental representation follow an area law.
In such theories, there is a phase transition between the confining phase and the Higgs phase, allowing us to clearly distinguish between them.
We will call these theories ``$t$-confining'', where ``$t$'' stands for ``truly.''
Note that a theory with fundamental matters like the supersymmetric version of the real-world quantum chromodynamics (QCD) is not a $t$-confining theory because its Wilson loops exhibit a perimeter law.
In this section, we start by defining what $t$-confinement means.

Specifically, we define $t$-confining theories as those that satisfy the following three conditions, which ensure the existence of an independent confinement phase.
Furthermore, in order to focus on local physics, we do not consider the differences in the coverings of the gauge group.
Therefore, unless otherwise stated, we will assume the universal covering group in the following discussions.
(For convenience, the $Spin$ group is sometimes referred to as the $SO$ group.)

The first condition is the unscreening condition.
This requires that the center of the gauge group remain unscreened and that Wilson lines sitting in the fundamental representation exhibit the area law.
If the center of the gauge group is screened, the confinement and Higgs phases connect smoothly, and it cannot be clearly distinguished between them \cite{PhysRevD.19.3682,BANKS1979349}.
Let $G$ be a gauge group and $\bar{G}$ be its universal covering group.
The center of $\bar{G}$ is denoted by $Z(\bar{G})=\{z\in\bar{G}| \forall g\in\bar{G}, zg=gz \}$.
To say that the center of the gauge group is not screened means that the identity element cannot be constructed from combinations of matter fields.
This implies that a non-trivial center remains:
\begin{equation}
\label{Eq:center}
    Z(\bar{G})/\text{gcd}(q_1,q_2,\cdots,q_m)\neq1
\end{equation}
where $q_1,q_2,\cdots,q_m$ are the charges of the matter fields under the center symmetry and gcd is greatest common divisor.
The center charge corresponds to the number of boxes in the Young tableau (see appendix \ref{sec:data_of_Lie_algebra}).
For example, in the case of the $SU(n)$ group, the center charge is $1$ for the fundamental representation and $n$ for the adjoint representation.
Wilson line operators
\begin{equation}
    \mathcal{W}_{R}(L)=\tr_{R}\mathcal{P}\exp\Big(2\pi i\int_{L}A\Big)
\end{equation}
are charged under the center symmetry.
Here, $R$ denotes an representation of the gauge group, $L$ is a line in spacetime, $\mathcal{P}$ denotes path ordering, and $A$ is the gauge field.
If the starting and ending points are defined to be equal, the line $L$ becomes a loop, the operator becomes gauge-invariant, and the vacuum expectation value in the large loop limit can be used to distinguish different phases of the gauge theory.
Of course, its behavior reflects the condition \ref{Eq:center}.
If that remaining center symmetry is nontrivial, a linear potential between charged particles and an area law for the Wilson line are expected.

The second condition is the index condition.
We impose the following inequality involving the Dynkin indices of the gauge and matter fields:
\begin{equation}
\label{Eq:index_condition}
    \sum_{i}I_i<I_G +2
\end{equation}
where $I_i$ are the Dynkin indices of the matter fields and $I_G$ is the Dynkin index of the gauge field. 
By imposing this Dynkin index constraint on a theory, we can extract those theories that exhibit a confining phase from among the vast number of possible theories.
This significantly narrows the scope of our investigation.
This method has often been used in the systematic classification of theories (see, for example, \cite{Csaki:1996sm,Csaki:1996zb,Csaki:1998dp}).
However, note that unlike previous studies, we do not derive the Dynkin index condition using $R$-symmetry.  
Based on empirical observations, when the index condition is reversed as $\sum_{i}I_i\ge I_G +2$, the theory flows to an $s$-confining, IR-free, or non-Abelian Coulomb phase and is not expected to $t$-confine.

The third condition is 't Hooft anomaly matching conditions \cite{tHooft:1979rat}.
When considering confining phases, it is expected that the UV and IR anomalies match.
It is also important to require the anomaly matching conditions for discrete symmetries such as ${\mathbb Z}_N$ \cite{Csaki:1997aw}. For (discrete)(gravity)$^2$ anomalies, they are supposed to match modulo $N/2$, given that a gravitational instantons on spin manifolds have two zero modes. Other anomalies ${\mathbb Z}_N {\rm gauge}^2$, ${\mathbb Z}_N^2 U(1)$, ${\mathbb Z}_N^3$, need to match modulo $N$, assuming no charge fractionalization.

If an anomaly associated with discrete symmetries does not match, it implies that the discrete symmetries are spontaneously broken, which may indicate the condensation of some ``magnetic'' degrees of freedom. In supersymmetric gauge theories, there are moduli space of vacua. If the origin is included in the moduli space, it is most appropriate to compute the anomalies at the origin, where the conditions are the most stringent. In the following analysis, we maintain this approach by identifying the branch that includes the origin within the moduli space, even in the case of multi-branch theories we encounter.

We define the $t$-confining theory by the three conditions above.
From here, we classify the $t$-confining theory and perform an additional analysis on models that have not yet been studied.

\section{Classification}
\label{sec:classification}

In this section, we classify the supersymmetric $t$-confining theories defined above.
Throughout this paper, we focus on theories that have a simple gauge group and no tree-level superpotential.
Now, we begin by listing the theories that satisfy the first two of the three conditions defining $t$-confining theories, as introduced before.
That is, we examine theories that satisfy the index condition \ref{Eq:index_condition} without including matter fields in the fundamental or spinor representations.
\footnote{
For $SO(4k)$, the center group is ${\mathbb Z}_2\times{\mathbb Z}_2$.
There are two spinor representations which are trivial under one of the ${\mathbb Z}_2$ factor.
That is, one has the $(1,0)$ charge under $\bZ_2\times\bZ_2$, while the other has the $(0,1)$ charge.
Therefore note that even if the theory contains a single spinor, it does not necessarily mean that all center symmetris are screened.
This point is taken into account in our classification.
}
Since matter fields in large representations of the gauge group tend to have large Dynkin indices, satisfying the index condition becomes challenging.
At this stage, we find that the number of candidate theories is significantly reduced.
Naively, the theories that satisfy the two conditions are as follows.
\renewcommand{\labelenumi}{(\arabic{enumi})}
\begin{enumerate}
\item $G$ : $\mathcal{N}=1$ Yang-Mills 
\item $SU(N_c)+\ydiagram{1,1}+\overline{\ydiagram{1,1}}$
\item $SU(4)+N_f\,\ydiagram{1,1}\,,\,\,\,N_f=1,2,3,4$
\item $SU(6)+N_f\,\ydiagram{1,1,1}\,,\,\,\,N_f=1,2$
\item $Sp(N_c)+\ydiagram{1,1}$
\item $Sp(2)+N_f\,\ydiagram{1,1}\,,\,\,\,N_f=1,2,3$
\item $Sp(3)+N_f\,\ydiagram{1,1}\,,\,\,\,N_f=1,2$
\item $Sp(3)+\ydiagram{1,1}+\ydiagram{1,1,1}$
\item $Sp(3)+\ydiagram{1,1,1}$
\item $SO(N_c)+N_f\,\ydiagram{1}\,,\,\,\,N_f\le N_c-2$
\item $SO(12)+N_f\,S\,,\,\,\,N_f=1,2$
\item $G+\adj$ : $\mathcal{N}=2$ Yang-Mills
\end{enumerate}
Note that not all of these theories are $t$-confining theories.
(2) does not have a center symmetry when $N_c$ is odd.
(3) is locally equivalent to $SO(6)+N_f\,\ydiagram{1}$, and has no vacuum for $N_f=1$ and has an Abelian Coulomb phase for $N_f=4$.
(4) has an Abelian Coulomb phase for $N_f=2$ \cite{Csaki:1998dp}.
(6) is locally equivalent to $SO(5)+N_f\,\ydiagram{1}$, and has an Abelian Coulomb phase for $N_f=3$.
(7) has an Abelian Coulomb phase for $N_f=2$ \cite{Csaki:1998dp}.
(8) and (9) do not have a center symmetry.
(10) has no vacuum for $N_f\le N_c-5$, and becomes an Abelian Coulomb phase for $N_f=N_c-2$ \cite{Intriligator:1995id}.
(11) has no vacuum for $N_f=1$.
Note that $S$ exhibits the matter field in the spinor representation.
(12) are $\mathcal{N}=2$ pure Yang-Mills theories that exhibit Abelian Coulomb phases for any Lie group $G$.
Note that we use the normalization that the fundamental representaion of $Sp(k)$ has dimension $2k$.
From the above discussion, we conclude that the theories with a nontrivial center symmetry and a confining phase are:
\renewcommand{\labelenumi}{(\Alph{enumi})}
\begin{enumerate}
    \item $G$
    \item $SO(k)+(k-4)\,\ydiagram{1}$
    \item $SO(k)+(k-3)\,\ydiagram{1}$
    \item $SU(6)+\ydiagram{1,1,1}$
    \item $Sp(k)+\ydiagram{1,1}$
    \item $SU(2k)+\ydiagram{1,1}+\overline{\ydiagram{1,1}}$
    \item $SO(12)+2\,S$
    
\end{enumerate}
Note that the order has been rearranged.
Now we need to examine whether these theories satisfy the third condition, 't Hooft anomaly matching conditions including discrete symmetries.
There are previous studies except for cases (E), (F) and (G).
(D) has been studied in \cite{Henning:2021ctv,Csaki:1997aw}, and (B) and (C) have been studied in \cite{Intriligator:1995id,Csaki:1997aw}.
In this paper, we first review cases from (A) to (D).
Furthermore, we analyze the dynamics of new models: (E) in sec. \ref{sec:Sp(2k)+A}, (F) in sec. \ref{sec:SU(2k)+A+A} and (G) in sec. \ref{sec:SO(12)+2s}.
This completes the entire analysis.

From this point onward, the corresponding Young diagram is sometimes identified based on the number of indices.
Specifically, matter fields in the vector representation, the antisymmetric representation (and its complex conjugate), and the rank three antisymmetric tensor representation are denoted by simplified notations: $V^i$, $A^{ij}$ (or $\tilde{A}_{ij}$), and $A^{ijk}$, respectively.

\section{Pure Yang-Mills}
\label{sec:pureYM}

The next four sections review previous research from the view point of $t$-confinement.
In the preceding sections, we classified $t$-confining theories.
Among them, two series of models, those with gauge groups $SU(2k)$ and $Sp(k)$ remain unanalyzed.
Before diving into these, we begin by examining simpler models.
We start with the pure Yang-Mills theory, followed by $SO$ gauge theories with vector matters, and then $SU(6)+A^{ijk}$ theory.
Readers already familiar with these models can skip ahead to the sec. \ref{sec:Sp(2k)+A}.

Pure Yang-Mills theory is believed to exhibit confinement with a mass gap as a result of gaugino condensation.
Its low energy effective theory is described by a glueball superfield with a Veneziano--Yankielowicz superpotential.
After integrating out all dynamical degrees of freedom, the superpotential is uniquely determined by symmetry and holomorphy as
\begin{equation}
    W_{\text{eff}} = c \Lambda^3
\end{equation}
where instanton calculations have shown that the constant $c$ is non-zero \cite{DAVIES1999123,NOVIKOV1985157,VZS85}.
If the complexified gauge coupling $\tau$ is regarded as a chiral superfield, then the source of the scalar component $\tr\lambda\lambda$ of the kinetic term of the gauge $\tr \cW^\alpha \cW_\alpha$ is given by the auxiliary field component $F_\tau$ of $\tau$.
Accordingly, the vacuum expectation value of $\tr\lambda\lambda$ is given by:
\begin{equation}
    \langle \operatorname{tr} \lambda \lambda \rangle = 16\pi i \frac{\partial S_{\text{eff}}}{\partial F_\tau} = 16\pi i \frac{\partial W_{\text{eff}}}{\partial \tau} \propto \Lambda^3 = (\Lambda^{3h})^{1/h}
\end{equation}
where $h=I_G/2$ is the dual Coxeter number, or equivalently a half Dynkin index of the adjoint representation of the gauge group.
Since $\Lambda^{3h}$ is single-valued, it follows that the supersymmetric Yang-Mills theory has $h$ branches. It agrees with the number of vacua computed by the Witten index for classical groups \cite{Witten:1982df,Witten:1997bs}.

Pure Yang-Mills theory contains no matter fields, so it possesses a nontrivial center symmetry.
Since the Wilson loop exhibits an area law, the theory satisfies the first condition for being a $t$-confining theory.
The second condition is trivially satisfied because of the absence of matter fields.
Before verifying the third condition, let us review the charges of fields in pure Yang-Mills theory.
The theory has a mass gap, so no gauge-invariant fields appear in the low-energy spectrum and are therefore not listed.
\begin{table}[H]
    \centering
    \myspace{1.5}
    \begin{tabular}{ScScScScSc}
        \hline
        symmetry & $G$ & $\mathbb{Z}_{2h}$ \\
        \hline
        $\mathcal{W}^\alpha$\rule[-1mm]{0mm}{5mm} & $\adj$ & $1$ \\
        \hline
    \end{tabular}
    \caption{pure $G$ Yang-Mills}
\end{table}
\noindent
The $U(1)_R$ symmetry is expected to be broken to $\mathbb{Z}_{2h}$ due to instanton effects.
This is reflected in the table.
However, the gaugino condensate breaks the symmetry further down to $\mathbb{Z}_2$, a subgroup of the Lorentz group.
This fact can also be confirmed by imposing the anomaly matching condition for the discrete symmetry.
\begin{table}[H]
    \centering
    \renewcommand{\arraystretch}{1.5}
    \begin{tabular}{l|l|l|l}
        anomaly & $\mathrm{UV}$ & $\mathrm{IR}$ & $\mathrm{UV}-\mathrm{IR}$ \\
        \hline
        $\tr\bZ_{2h}\mathrm{(gravity)}^2$ & $1\times\dim(G)$ & $0$ & $\dim(G)$ \\
        $\tr\bZ_{2h}^3$ & $1^3\times\dim(G)$ & $0$ & $\dim(G)$ \\
    \end{tabular}
    \caption{anomaly matching conditions for pure Yang-Mills}
\end{table}

\noindent
From this, it follows that the anomalies should match modulo $1$.
This provides an alternative way to confirm that the $R$-symmetry is broken down to $\mathbb{Z}_2$.

Similar phenomena will appear in many of the $t$-confining theories we will examine later.
In such cases, the condensation of glueball fields, typically involving hybrid ``magnetic'' operators of quarks and gluons, results in the breaking of discrete symmetries.  
Therefore, determining the modulo class in which the discrete anomaly matching condition is satisfied becomes critically important.

\section{$SO(k)$ with $k-4$ $V^i$s}
\label{sec:SO(N)+(N-4)V}

Next example is an $SO(k)$ gauge theory with vectors studied by Intriligator and Seiberg \cite{Intriligator:1995id}.
Let us begin by briefly reviewing this theory.
The squark vacuum expectation value breaks the gauge group down to $SO(4) = SU(2) \times SU(2)/\mathbb{Z}_2$, and gaugino condensation in each $SU(2)$ factor generates a superpotential:
\begin{equation}
    W=\frac{1}{2}(\epsilon_L+\epsilon_R)\Big(\frac{16\Lambda^{2(k-1)}}{\det M}\Big)^{1/2}
\end{equation}
where $\epsilon_L,\epsilon_R=\pm1$.
There are two physically equivalent branches, $\epsilon_L\epsilon_R=1$ and $\epsilon_L\epsilon_R=-1$.
In the latter branch, the superpotential vanishes: $W=0$.
Furthermore, since the anomaly matching conditions for the continuous symmetries are satisfied at the origin of the moduli space, $M^{\{ij\}}$ is considered to describe low energy physics.
Note that although the chiral symmetry is unbroken at the origin, there is no complementarity, so the theory is not $s$-confining.
In particular, note that by definition, $s$-confinement and $t$-confinement are mutually exclusive.

Now, let us check whether this theory is $t$-confining.
First, it is immediately clear that the $\mathbb{Z}_2$ center is not screened, since there are no matter fields in the spinor representation.
A straightforward calculation also shows that the index condition is satisfied: $2(k-4)<2(k-2)+2$, where Dynkin index of the gauge field is $I(\adj)=2(k-2)$, and the sum of Dynkin indices of the matter fields is $(k-4)I(\ydiagram{1})=2(k-4)$.
Finally, let us check anomaly matching conditions at the origin of the moduli space of the $W=0$ branch including the discrete symmetry.
The charges of the fields contained in the theory are as follows:
\begin{table}[H]
    \centering
    \myspace{1.5}
    \begin{tabular}{ScScScScSc}
        \hline
        symmetry & $SO(k)$ & $SU(k-4)$ & $U(1)_R$ & $\bZ_{2k-8}$ \\
        \hline
        $\mathcal{W}^\alpha$\rule[-1mm]{0mm}{5mm} & $\adj$ & $\one$ & $1$ & $0$ \\
        $V^i$ & $\ydiagram{1}$ & $\ydiagram{1}$ & $-\dfrac{2}{k-4}$ & $1$ \\
        \hline
        $M^{\{ij\}}$ & $\one$ & $\ydiagram{2}$ & $-\dfrac{4}{k-4}$ & $2$ \\
        \hline
    \end{tabular}
    \caption{$SO(k)+(k-4)V^i$}
\end{table}
\noindent
The table above reflects the breaking of the $U(1)$ symmetry to $\mathbb{Z}_{2k-8}$ due to the instanton effect.
Based on this table, we can compute the anomaly matching conditions, including those for discrete symmetries.
The results are as follows:
\begin{table}[H]
    \centering
    \renewcommand{\arraystretch}{2}
    \tiny
    \begin{tabular}{l|l|l|l}
        anomaly & $\mathrm{UV}$ & $\mathrm{IR}$ & $\mathrm{UV}-\mathrm{IR}$ \\
        \hline
        $\tr U(1)_R(\mathrm{gravity})^2$ & $1\times\dfrac{k(k-1)}{2}+(-\dfrac{2}{k-4}-1)\times k\times(k-4)$ & $(-\dfrac{4}{k-4}-1)\times\dfrac{(k-4)(k-3)}{2}$ & $0$ \\
        $\tr SU(k-4)^3$ & $1\times k$ & $(k-4)+4$ & $0$ \\
        $\tr SU(k-4)^2U(1)_R$ & $1\times(-\dfrac{2}{k-4}-1)\times k$ & $\{(k-4)+2\}\times(-\dfrac{4}{k-4}-1)$ & $0$ \\
        $\tr U(1)_R^3$ & $(-\dfrac{2}{k-4}-1)^3\times k\times(k-4)+1^3\times\dfrac{k(k-1)}{2}$ & $(-\dfrac{4}{k-4}-1)^3\times\dfrac{(k-4)(k-3)}{2}$ & $0$ \\
        $\tr\bZ_{2k-8}\mathrm{(gravity)}^2$ & $1\times k\times(k-4)$ & $2\times\dfrac{(k-4)(k-3)}{2}$ & $3(k-4)$ \\
        $\tr\bZ_{2k-8}SU(k-4)^2$ & $1\times1\times k$ & $2\times\{(k-4)+2\}$ & $-(k-4)$ \\
        $\tr\bZ_{2k-8}U(1)_R^2$ & $1\times\{-2-(k-4)\}^2\times k\times(k-4)$ & $2\times\{-4-(k-4)\}^2\times\dfrac{(k-4)(k-3)}{2}$ & $-k(k-4)^2$ \\
        $\tr\bZ_{2k-8}^2U(1)_R$ & $1^2\times\{-2-(k-4)\}\times k\times(k-4)$ & $2^2\times\{-4-(k-4)\}\times\dfrac{(k-4)(k-3)}{2}$ & $k(k-4)^2$ \\
        $\tr\bZ_{2k-8}^3$ & $1^3\times k\times(k-4)$ & $2^3\times\dfrac{(k-4)(k-3)}{2}$ & $-3(k-4)^2$ \\
    \end{tabular}
    \caption{anomaly matching conditions for $SO(k)+(k-4)V^i$}
\end{table}
\noindent
Note that in the seventh and eighth rows of the table, the $R$-charges are normalized to be integers.
When $k$ is even, the $R$-charges can be further normalized by a factor of one-half.
The anomalies for the continuous symmetries match exactly.
However, the anomalies for the discrete symmetries do not.
To obtain the correct anomaly matching conditions, the discrete symmetry must be broken from $\mathbb{Z}_{2k-8}$ to $\mathbb{Z}_{k-4}$.
The $\mathbb{Z}_{k-4}$ symmetry is included in the $SU(k-4)$ flavor symmetry as a subgroup.
Therefore, if we check the anomaly matching relevant to $SU(k-4)$, there is no need to consider $\mathbb{Z}_{k-4}$.
The symmetry breaking is triggered by the condensation of the ``magnetic'' operator $\tr(V^{k-4}\cW^\alpha\cW_\alpha)$ \cite{Csaki:1997aw}.
This model also provides an example of a $t$-confining theory exhibiting condensation of the ``magnetic'' operator and the breaking of the discrete symmetry.

\section{$SO(k)$ with $k-3$ $V^i$s}
\label{sec:SO(N)+(N-3)V}

Another well-known example is given by Intriligator and Seiberg \cite{Intriligator:1995id}.
In this theory, the dynamically generated superpotential is given by
\begin{equation}
    W=4(1+\epsilon)\frac{\Lambda^{2k-3}}{\det M}
\end{equation}
where $\epsilon=\pm1$.
This superpotential reflects the instanton effect and gaugino condensation when the gauge symmetry is broken to $SO(3)$.
For $\epsilon=-1$, there is a branch where $W = 0$, which is realized effectively.
We discuss a superpotential with exotic baryons later.
Therefore, we can again check the anomaly matching conditions at the origin.

Now, we can easily check whether or not this theory is $t$-confining.
First, since this theory does not contain matter fields in the spinor representation, the center is not screened.
Next, it is also easy to see that the index condition $2(k-3)<2(k-2)+2$ is satisfied.
Finally, we check the anomaly matching conditions.
The charges of the fields contained in the theory are as follows:
\begin{table}[H]
    \centering
    \myspace{1.5}
    \begin{tabular}{ScScScScSc}
        \hline
        symmetry & $SO(k)$ & $SU(k-3)$ & $U(1)_R$ & $\bZ_{2k-6}$ \\
        \hline
        $\mathcal{W}^\alpha$\rule[-1mm]{0mm}{5mm} & $\adj$ & $\one$ & $1$ & $0$ \\
        $V^i$ & $\ydiagram{1}$ & $\ydiagram{1}$ & $-\dfrac{1}{k-3}$ & $1$ \\
        \hline
        $M^{\{ij\}}$ & $\one$ & $\ydiagram{2}$ & $-\dfrac{2}{k-3}$ & $2$ \\
        $q_i$ & $\one$ & $\overline{\ydiagram{1}}$ & $\dfrac{k-2}{k-3}$ & $k-4$ \\
        \hline
    \end{tabular}
    \caption{$SO(k)+(k-3)V^i$}
\end{table}
\noindent
Note that when one flavor is decoupled, it is necessary to introduce a new massless field $q_i$ at the origin in order for the theory to correctly deform to a theory with $k - 4$ flavors.
In addition, the table above reflects the breaking of the $U(1)$ symmetry to $\mathbb{Z}_{2k-6}$ due to the instanton effect.
It can be confirmed that the anomaly matching conditions are fully satisfied.
\begin{table}[H]
    \centering
    \renewcommand{\arraystretch}{2}
    \tiny
    \begin{tabular}{p{2.2cm}|p{4.2cm}|p{4.2cm}|p{2.5cm}}
        anomaly & $\mathrm{UV}$ & $\mathrm{IR}$ & $\mathrm{UV}-\mathrm{IR}$ \\
        \hline
        $\tr U(1)_R(\mathrm{gravity})^2$ & $1\times\dfrac{k(k-1)}{2}+(-\dfrac{1}{k-3}-1)\times k\times(k-3)$ & $(-\dfrac{2}{k-3}-1)\times\dfrac{(k-3)(k-2)}{2}+(\dfrac{k-2}{k-3}-1)\times(k-3)$ & $0$ \\
        $\tr SU(k-3)^3$ & $1\times k$ & $\{(k-3)+4\}+(-1)$ & $0$ \\
        $\tr SU(k-3)^2U(1)_R$ & $1\times(-\dfrac{1}{k-3}-1)\times k$ & $\{(k-3)+2\}\times(-\dfrac{2}{k-3}-1)+1\times(\dfrac{k-2}{k-3}-1)$ & $0$ \\
        $\tr U(1)_R^3$ & $(-\dfrac{1}{k-3}-1)^3\times k\times(k-3)+1^3\times\dfrac{k(k-1)}{2}$ & $(-\dfrac{2}{k-3}-1)^3\times\dfrac{(k-3)(k-2)}{2}+(\dfrac{k-2}{k-3}-1)^3\times(k-3)$ & $0$ \\
        $\tr\bZ_{2k-6}\mathrm{(gravity)}^2$ & $1\times k\times(k-3)$ & $2\times\dfrac{(k-3)(k-2)}{2}+(k-4)\times(k-3)$ & $-(k-6)(k-3)$ \\
        $\tr\bZ_{2k-6}SU(k-3)^2$ & $1\times1\times k$ & $2\times\{(k-3)+2\}+(k-4)\times1$ & $-2(k-3)$ \\
        $\tr\bZ_{2k-6}U(1)_R^2$ & $1\times\{-1-(k-3)\}^2\times k\times(k-3)$ & $2\times\{-2-(k-3)\}^2\times\dfrac{(k-3)(k-2)}{2}+(k-4)\times\{(k-2)-(k-3)\}^2\times(k-3)$ & $-2(k-3)^2$ \\
        $\tr\bZ_{2k-6}^2U(1)_R$ & $1^2\times\{-1-(k-3)\}\times k\times(k-3)$ & $2^2\times\{-2-(k-3)\}\times\dfrac{(k-3)(k-2)}{2}+(k-4)^2\times\{(k-2)-(k-3)\}\times(k-3)$ & $-4(k-3)^2$ \\
        $\tr\bZ_{2k-6}^3$ & $1^3\times k\times(k-3)$ & $2^3\times\dfrac{(k-3)(k-2)}{2}+(k-4)^3\times(k-3)$ & $-(k-3)^2(k^2-9k+24)$ \\
    \end{tabular}
    \caption{anomaly matching conditions for $SO(k)+(k-3)V^i$}
\end{table}
\noindent
The spontaneous breaking of discrete symmetries observed in the case with $k-4$ flavors is not seen here.
Hence, it is expected that no ``magnetic'' operator condensates.
It will be revealed later that this model is the only one in $t$-confining theories where no condensate exists.

However, there is an advantage that was absent in the $k-4$ case: the theory with $k-3$ flavors arises as a branch where magnetic monopoles in the theory with $k-2$ flavors condense, thus exhibiting the dual Meissner effect.
Namely that the theory with $k-3$ flavors has a Coulomb branch parameterized by $u = {\rm det}M$. There are two singularities but we focus on one at $u=0$. There are massless monopoles with a flavor index $E^\pm_i$ at this singularity with a superpotential
\begin{align}
    W = \frac{1}{\mu} M^{ij} E^{+}_i E^{-}_j,
\end{align}
where $i,j = 1, \cdots, k-3$. 
Upon perturbation with a mass for the $k-3$-th flavor, $E^{\pm}_{k-3}$ condense, breaking the U(1) gauge group,  causing the dual Meissner effect and hence confinement. The resulting theory with $k-4$ flavors has the superpotential
\begin{align}
    W = \frac{1}{\mu} M^{ij} q_i q_j.
\end{align}
where $i,j = 1, \cdots, k-4$, and $q_i$ is identified with exotic baryons 
\begin{align}
    q_i = \epsilon_{i j_1 \cdots j_{k-4}}
    \epsilon^{r_1 \cdots r_{k-4} r_{k-3} r_{k-2} r_{k-1} r_k}
    V^{j_1}_{r_1} \cdots V^{j_{k-4}}_{r_{k-4}} {\cal W}^\alpha_{r_{k-3} r_{k-2}} {\cal W}^\beta_{r_{k-1} r_{k}} \epsilon_{\alpha\beta}. 
\end{align}
Here, $i, j_n$ are $SU(k-3)$ flavor indices, $r_n$ are $SO(k)$ gauge indices, and $\alpha,\beta=1,2$ are spinor indices. Even though $W\neq 0$, the equation of motion with respect to $M^{ij}$ requires $\partial W/\partial M^{ij} = q_i q_j = 0$, and hence $q_i = 0$. Therefore the superpotential effectively vanishes.

Note that there is a fundamental difference from the theory with $k-4$ flavors.
The condensate here is not something that can be defined within the original theory, but rather a magnetic monopole obtained from an upstream theory with $k-2$ flavors.
If we consider the condensation of ``magnetic'' objects in a $t$-confining theory to include operators from not only the original theory but also from upstream theories, then we can say that condensates appear in all $t$-confining theories.

\section{$SU(6)$ with $A^{ijk}$}
\label{sec:SU(6)+A}

The theory with gauge group $SU(6)$ and a rank-3 antisymmetric tensor field is also a $t$-confining theory.
This theory was analyzed in \cite{Henning:2021ctv}, but it was not emphasized that the Wilson loops exhibit an area law indicating strict confinement.
Since the matter field in this theory belongs to a pseudo-real representation of the gauge group, this theory is chiral.
It is noteworthy that this is the only one chiral gauge theory among the $t$-confining theories.

First, we briefly review the dynamics of the theory following \cite{Henning:2021ctv}.
The field content of this theory is as follows:
\begin{table}[H]
    \centering
    \myspace{1.5}
    \begin{tabular}{ScScScSc}
        \hline
        symmetry & $SU(6)$ & $U(1)_R$ & $\bZ_6$ \\
        \hline
        $\mathcal{W}^\alpha$\rule[-1mm]{0mm}{5mm} & $\adj$ & $1$ & $0$ \\
        $A^{ijk}$ & $\ydiagram{1,1,1}$ & $-1$ & $1$ \\
        \hline
        $A^4$ & $\one$ & $-4$ & $4$ \\
        \hline
    \end{tabular}
    \caption{$SU(6)+A^{ijk}$}
\end{table}
\noindent
Since $A^{ijk}$ transforms in a pseudo-real representation, there is no invariant of the form $A^2$, and the lowest-order invariant is $A^4$.
Along the D-flat direction, the gauge group is spontaneously broken to $SU(3) \times SU(3)$.
Each $SU(3)$ factor undergoes gaugino condensation, and the resulting superpotential is:
\begin{equation}
    W=(\omega^{n_1}-\omega^{n_2})\frac{\Lambda^5}{(A^{4})^{1/2}}
\end{equation}
where $\omega=e^{2\pi i/3}$ and $n_1, n_2=0,1,2$.
It is found that a branch where $W = 0$ appears again when $n_1 = n_2$.
In fact, the branch where the superpotential vanishes can be observed in all $t$-confining theories, except for pure Yang-Mills.
When checking the anomaly matching conditions at the $W = 0$ branch, it is found that the anomaly of $Z_6$ does not match:
\begin{table}[H]
    \centering
    \renewcommand{\arraystretch}{1.7}
    \small
    \begin{tabular}{p{3cm}|p{6cm}|p{2cm}|p{1.2cm}}
        anomaly & $\mathrm{UV}$ & $\mathrm{IR}$ & $\mathrm{UV}-\mathrm{IR}$ \\
        \hline
        $\tr U(1)_R(\mathrm{gravity})^2$ & $1\times(6^2-1)+(-1-1)\times\dfrac{6\cdot5\cdot4}{6}$ & $-4-1$ & $0$ \\
        $\tr U(1)_R^3$ & $1^3\times(6^2-1)+(-1-1)^3\times\dfrac{6\cdot5\cdot4}{6}$ & $(-4-1)^3$ & $0$ \\
        $\tr\bZ_6\mathrm{(gravity)}^2$ & $1^3\times\dfrac{6\cdot5\cdot4}{6}$ & $4$ & $16$ \\
        $\tr\bZ_6U(1)_R^2$ & $1(-1-1)^2\times\dfrac{6\cdot5\cdot4}{6}$ & $4(-4-1)^2$ & $-20$ \\
        $\tr\bZ_6^2U(1)_R$ & $1^2(-1-1)\times\dfrac{6\cdot5\cdot4}{6}$ & $4^2(-4-1)$ & $40$ \\
        $\tr\bZ_6^3$ & $1^3\times\dfrac{6\cdot5\cdot4}{6}$ & $4^3$ & $-44$ \\
    \end{tabular}
    \caption{anomaly matching conditions for $SU(6)+A^{ijk}$}
\end{table}
\noindent
As we observed in previous $t$-confining theories, this is caused by the condensation of a ``magnetic'' operator.
In this case, $\tr(A^2\cW_\alpha\cW^\alpha)$ condenses, which causes the discrete symmetry breaking from $Z_6$ to $Z_2$, and the anomaly matching conditions are confirmed under modulo $2$.

Now, we confirm that this theory is a $t$-confining theory.
The center of the $SU(6)$ gauge group is $\bZ_6$, but screening occurs due to the presence of $A^{ijk}$.
Since the center charge of $A^{ijk}$ is $3$, the subgroup $\bZ_3$ remains unscreened.
Thus, there exist Wilson line operators that exhibits the area law, satisfying the first condition of $t$-confinement.
The Dynkin index of the gauge field is $I(\adj)=2\times6=12$, and the Dynkin index of the matter field is $I(A^{ijk})=(6-3)(6-2)/2=6$.
Thus, the second condition of $t$-confining theories is satisfied: $6<12+2$.
Furthermore, the anomaly is completely matched due to the condensation of the ``magnetic'' operator $\tr(A^2\cW_\alpha\cW^\alpha)$, which leads to the spontaneous breaking of the discrete symmetry $\bZ_6\to\bZ_2$ \cite{Henning:2021ctv,Csaki:1997aw}.
As a result, it is confirmed that the $SU(6)+A^{ijk}$ theory indeed satisfies the conditions for $t$-confinement.

\section{$Sp(k)$ with $A^{ij}$}
\label{sec:Sp(2k)+A}

From this point onward, we will explore new $t$-confining models.

In this section, we analyze the $Sp(k)+A^{ij}$ gauge theory.
We adopt a convention in which the dimension of the fundamental representation is $2k$.
This model was previously analyzed in reference \cite{Cho:1996bi,Csaki:1996eu}, but the analysis was incomplete.
In our work, we newly perform calculations of the superpotential for general $k$, derive the existence of $W = 0$ branches, study the condensation of a ``magnetic'' operator, examine anomaly matching conditions for discrete symmetries, and check various consistency conditions.

First, let us explain the basic structure of this theory.
We denote the invariant tensor of the symplectic group by $J=\mathrm{1}_{k\times k}\otimes i\sigma_2$, where $\sigma_2$ is a Pauli matrix.
Note that the antisymmetric representation should not contain the trivial representation.
Thus, the following condition holds $\tr(AJ)=0$.
This is a consequence of the irreducibility of the representation.
The charges of the fields in the theory are as follows:
\begin{table}[H]
    \centering
    \myspace{1.5}
    \begin{tabular}{ScScScSc}
        \hline
        symmetry & $Sp(k)$ & $U(1)_R$ & $\bZ_{2k-2}$ \\
        \hline
        $\cW^\alpha$\rule[-1mm]{0mm}{5mm} & $\adj$ & $1$ & $0$ \\
        $A^{ij}$ & $\ydiagram{1,1}$ & $-\dfrac{2}{k-1}$ & $1$ \\
        \hline
        $\tr(AJ)^i$ & $\one$ & $-\dfrac{2i}{k-1}$ & $i$ \\
        \hline
    \end{tabular}
    \caption{$Sp(k)+\ydiagram{1,1}$}
\end{table}
\noindent
The $U(1)$ symmetry is expected to break to $\bZ_{I(A^{ij})}=\bZ_{2k-2}$ due to instantons.
However, it later becomes clear that the discrete symmetry is actually broken down to $\mathbb{Z}_{k-1}$.
The one loop coefficient of the beta function is given by $b=3(k+1)-(k-1)=2k+4$.
The classical moduli space is spanned by the following eigenvalues:
\begin{equation}
\label{Eq:moduli_Sp_asym}
    A=
    \begin{pmatrix}
        a_1 && \\
        & \ddots & \\
        && a_n
    \end{pmatrix}
    \otimes\sigma_2
\end{equation}
where, due to the irreducibility of the antisymmetric representation of $Sp(k)$, the trace-less condition must be satisfied:
\begin{equation}
    \sum_ia_i=0
\end{equation}
Therefore, we see that at a generic point of the moduli space, the gauge group is broken to $SU(2)^k$.
The classical moduli space is also expressed in gauge invariant quantities, that is
\begin{equation}
    \tr(AJ)^i
\end{equation}
where $i=2,\cdots,k$.
There are no relations between the gauge invariant operators.
Since the Pfaffian $\Pf A$ is not an independent invariant, it is omitted.
Through the super Higgs mechanism,\footnote{
Note that the term ``super-Higgs mechanism'' has two distinct meanings in supersymmetric gauge theories \cite{Terning:2006bq}. One refers to the case where a vector superfield eats a goldstone field, and the other refers to the case where a Rarita-Schwinger field eats a goldstino field.
In this context, we are using the former meaning.
}
$k(2k+1)-3k=2k^2-2k$ massless chiral superfields are eaten by massless vector superfields.
Therefore, the dimension of the classical moduli space is $\{k(2k-1)-1\}-(2k^2-2k)=k-1$, which agrees with the dimension found from the D-flatness condition and those obtained from the number of gauge-invariant operators.
We first perform calculations of the superpotential for general $k$ and check various consistency conditions.

\subsection{Superpotential}

Since the gauge group breaks to $SU(2)^k$ at general points on the moduli space, the superpotential of the theory should reflect all gaugino condensates.
Gaugino condensation contributes from each $SU(2)_i$ factor and is given by:
\begin{equation}
\label{Eq:Sp_gaugino}
    W_\mathrm{eff}=\sum_i\pm(\Lambda_i^6)^{\frac{1}{2}}
\end{equation}
The sign choices for each factor lead to multiple branches.
We do not immediately derive the superpotential of the theory from this point.
First, we conjecture a superpotential that reproduces the above expression when the vector superfield is integrated.
Fortunately, in supersymmetric gauge theories, the constraints from symmetry and holomorphy are very strong, allowing us to determine nearly all of the details of the specific form.
The expected form is as follows:
\begin{equation}
\label{Eq:Sp_superpotential}
    W_\mathrm{eff}=\sum_i\epsilon_i\left(\dfrac{\Lambda^{2k+4}}{\prod_{j\neq i}(a_i-a_j)^2}\right)^{\frac{1}{2}}
    =\sum_i\epsilon_i\dfrac{\Lambda^{k+2}}{\prod_{j\neq i}(a_i-a_j)}
\end{equation}
where $\epsilon_i=\pm1$.
Note that, from the action of the Weyl group, permutations of $a_i$ are gauge equivalent, and from the $R$-symmetry of $W_\mathrm{eff}$, flipping the signs of $\epsilon_i$ is also an equivalent transformation.
For the action of the Weyl group, please refer to the appendix \ref{sec:actions_of_the _Weyl_groups}.
This results in a theory with many branches.
In particular, if all the signs of $\epsilon_i$ are the same, a branch with $W_\mathrm{eff}=0$ arises.
A detailed proof of this fact is provided in the appendix \ref{sec:W=0}.
The consistency of this superpotential can be confirmed in several ways.

\paragraph{Super Higgs Mechanism}

By performing a dynamical scale matching via the super Higgs mechanism, it is possible to interpolate between \ref{Eq:Sp_gaugino} and \ref{Eq:Sp_superpotential}, and the superpotential can be derived.
In fact, by calculating the fermion mass from the Kähler potential $\int d^2\theta d^2\bar{\theta}\tr\{A(e^V)^\top A^\dagger e^V\}$, the holomorphic mass of the vector multiplet that acquires mass via the super Higgs mechanism can be computed.
When the gauge symmetry breaks, the holomorphic mass of the vector multiplet that becomes the fundamental representation of $SU(2)_i$ and the anti-fundamental representation of $SU(2)_j$ is given by $m_{i,j}=a_i-a_j$.
Thus, the dynamical scale matching for each $SU(2)_i$ factor is given by
\begin{equation}
    \dfrac{\Lambda^{2k+4}}{\prod_{j\neq i}m_{i,j}^2}=\dfrac{\Lambda^{2k+4}}{\prod_{j\neq i}(a_i-a_j)^2}=\Lambda_i^6
\end{equation}
This fact clearly holds for any arbitrary $i$.
Therefore, we can conclude that our superpotential \ref{Eq:Sp_superpotential} is correct.

\paragraph{Decoupling Colors}

Here, we demonstrate that by giving large vacuum expectation values to $a_{k-1}$ and $a_k$ in \ref{Eq:moduli_Sp_asym}, we can consistently transition to the smaller theory $Sp(k-2)+A^{ij}$.
Note, however, that due to the irreducibility of the antisymmetric representation, the sum of all eigenvalues must vanish, implying that $a_k=-a_{k-1}\to\infty$.
The dynamical scale matching condition is 
\begin{equation}
    \bigg(\frac{\Lambda}{a_k}\bigg)^b=\bigg(\frac{\Lambda_L}{a_k}\bigg)^{b_L}
\end{equation}
where $\Lambda_L$ is the dynamical scale in the low-energy $Sp(k-2)$ theory, and $b_L$ is the one loop coefficient of its beta function.
By easy computations, they are identified as $b=2k+4$ and $b_L=2k$, so we have $\Lambda^{2k+4}=\Lambda_L^{2k}{a_k}^4$.
When the eigenvalues have large vacuum expectation values, the superpotential becomes
\begin{align}
    W_\mathrm{eff}
    & =\sum_{i=1}^k\epsilon_i\dfrac{\Lambda^{k+2}}{\prod_{j\neq i}(a_i-a_j)} \\
    & =\sum_{i=1}^{k-2}\epsilon_i\dfrac{\Lambda^{k+2}}{(a_i-a_{k-1})(a_i-a_k)\prod_{j\neq i,k-1,k}(a_i-a_j)} \notag\\
    & \qquad\qquad\qquad+\epsilon_{k-1}\dfrac{\Lambda^{k+2}}{\prod_{j\neq k-1}(a_{k-1}-a_j)}+\epsilon_k\dfrac{\Lambda^{k+2}}{\prod_{j\neq k}(a_k-a_j)} \\
    & \to\sum_{i=1}^{k-2}\epsilon_i\dfrac{\Lambda^{k+2}}{(-a_{k-1})(-a_k)\prod_{j\neq i,k-1,k}(a_i-a_j)} \\
    & =\sum_{i=1}^{k-2}\epsilon_i\dfrac{\Lambda_L^{(k-2)+2}}{\prod_{j\neq i}(a_i-a_j)}
\end{align}
where we considered the limit $a_k \to \infty$ in the third line.
This result corresponds to the case of the $Sp(k-2)$ group, that is $Sp(k-2)+A^{ij}$. Namely the result passes the consistency check.

\paragraph{Decoupling Flavors}

Here, we analyze the behavior of the theory when quark masses are introduced.
We confirm that the number of vacua emerging after integrating out the quarks coincides with the $k+1$ vacua of the pure Yang-Mills theory.
Our analysis focuses on several small values of $k$.
The effective superpotential, which includes both mass terms and Lagrange multiplier terms, is given by
\begin{equation}
    W_\mathrm{eff}=2\Lambda^{k+2}\sum_i\dfrac{\epsilon_i}{\prod_{j\neq i}(a_i-a_j)}+\dfrac{1}{2}m\sum_ia_i^2+\lambda\sum_ia_i
\end{equation}
The vacuum solutions are given by solving the system of equations:
\begin{equation}
\label{Eq:sol_sp}
    \dfrac{\partial W_\mathrm{eff}}{\partial a_i}=\dfrac{\partial W_\mathrm{eff}}{\partial\lambda}=0\,,\quad(i=1,\cdots,k)
\end{equation}
The dynamical scale matching condition is $\Lambda_\mathrm{YM}^{3k+3}=m^{k-1}\Lambda^{2k+4}$, and the gaugino condensation in pure Yang-Mills theory become $W_\mathrm{eff}=(k+1)\Lambda_\mathrm{YM}^3=(k+1)(m^{k-1}\Lambda^{2k+4})^{\frac{1}{k+1}}$.
Any vacuum solution obtained by \ref{Eq:sol_sp} that does not reproduce the correct gaugino condensate corresponds to the case where more than one of the eigenvalues become the same, and therefore the gauge group does not break down to $SU(2)^k$.
Moreover, we do not distinguish vacua that are identified under the action of the Weyl group of the gauge group and the $R$-symmetry.

Under this approach, we can indeed confirm that the same number of vacua as in the Yang-Mills theory is generated.
For example, for $k=2$, it is sufficient to check for the case $\epsilon_1=-\epsilon_2=1$.
In this case, since the Yang-Mills theory has three vacua, we expect that \ref{Eq:sol_sp} also has three solutions.
Indeed, there are three solutions, which are given by:
\begin{equation*}
    (a_1,a_2)=\left(-\left(\dfrac{\Lambda^4}{m}\right)^\frac{1}{3},\left(\dfrac{\Lambda^4}{m}\right)^\frac{1}{3}\right),
        \left(\omega_3\left(\dfrac{\Lambda^4}{m}\right)^\frac{1}{3},-\omega_3\left(\dfrac{\Lambda^4}{m}\right)^\frac{1}{3}\right),
        \left(-\omega_3^2\left(\dfrac{\Lambda^4}{m}\right)^\frac{1}{3},\omega_3^2\left(\dfrac{\Lambda^4}{m}\right)^\frac{1}{3}\right)
\end{equation*}
These solutions reproduce the correct gaugino condensation in the pure Yang-Mills theory and give the correct Witten index.
For other cases with larger $k$, we can also obtain the correct Yang-Mills vacua using the same calculation method.
However, as $k$ increases, the choice of the sign $\epsilon_i$s for obtaining the Yang-Mills vacua becomes more complicated and requires careful attention.
For $k=3$, it is sufficient to check the cases $\epsilon_1=\epsilon_2=-\epsilon_3=1$.
For $k=4$, solving the system of equations leads to the correct gaugino condensation only for the branching $\epsilon_1=\epsilon_2=-\epsilon_3=-\epsilon_4=1$.
For $k=5$, the branching $\epsilon_1=\epsilon_2=\epsilon_3=-\epsilon_4=-\epsilon_5=1$ is the only one that leads to the correct gaugino condensation.
The correct gaugino condensation occurs only when all eigenvalues are distinct.
We saw that for small values of $k$, the results are consistent with pure Yang-Mills theory.

\paragraph{Consistency with $SO(5)+V$}

In the case of $k=2$, since the theory is locally equivalent to the $SO(5)$ gauge theory with a vector $SO(5)+V$, we can verify its consistency by checking whether or not the results are consistent with those obtained in \cite{Intriligator:1995id}.
Note that there is the isomorphism of Lie algebra $\mathfrak{sp}(2)\simeq\mathfrak{so}(5)$, and the antisymmetric representation of $\mathfrak{sp}(2)$ is equivalent to the vector representation of $\mathfrak{so}(5)$.
Thus, this is the special case of $SO(k)+(k-4)V^i$ reviewed in sec. \ref{sec:SO(N)+(N-4)V}.

In fact, the equivalence of the superpotential is almost evident.
The superpotential in the $SO(5)+V$ case is given by
\begin{equation}
    W_\mathrm{eff}\propto\left(\dfrac{\Lambda^8}{M}\right)^\frac{1}{2}
\end{equation}
where the meson fields are given by $M=V^2$.
The superpotential in the $Sp(2)+A^{ij}$ case (see \ref{Eq:Sp_superpotential}) is given by
\begin{equation}
    W_\mathrm{eff}\propto\left(\dfrac{\Lambda^8}{(a_1-a_2)^2}\right)^{\frac{1}{2}}
\end{equation}
The moduli space of the $Sp(2)$ theory appears to be given by the two eigenvalues $a_1$ and $a_2$ (see \ref{Eq:moduli_Sp_asym}), but due to the irreducibility of the antisymmetric representation, the actual independent quantity is only one: $a_2=-a_1$.
Therefore, the denominator of the above superpotential reduces to $4a_1^2$, which corresponds to the case of $SO(5)$ theory.

\paragraph{Using $s$-confining Theories}

Now, we consider an extension to the $s$-confining theory $Sp(k)+A^{ij}+6Q^i$ by adding quarks in the fundamental representation.
We expect that by integrating out the quarks in the fundamental representation, we recover the original theory\cite{Cho:1996bi,Csaki:1996eu}.
Below, we verify whether our superpotential remains after integrating out all $Q^i$'s.
Gauge invariants are defined as follows:
\begin{subequations}
\begin{align}
    T_i&=\tr A^i \\
    M_i&=QA^iQ
\end{align}
\end{subequations}
Since $s$-confining theories have restrictions on their superpotential, the functional form of the superpotential is known \cite{Csaki:1996sm,Csaki:1996zb}.
In addition, as for $s$-confining theories, the constraints of the classical moduli space is reproduced by the equations of motion.
This fact uniquely determines the superpotential.
For $k=3$, the $s$-confining superpotential is given by
\begin{equation}
    W_\mathrm{eff}\propto T_2^2M_0^3+8T_3M_1M_0^2-12T_2M_0^2M_2+48M_0M_2^2+48M_1^2M_2
\end{equation}
By adding the mass term $mM_0$ for the quarks in the fundamental representation to the superpotential, and integrating them, we recover all branches of the superpotential of the original theory \cite{Cho:1996bi}.
The branching with $W\neq0$ is given by
\begin{equation}
    W_\mathrm{eff}\propto\dfrac{1}{T_2\big\{(\sqrt{R}+\sqrt{R+1})^{\frac{2}{3}}+(\sqrt{R}+\sqrt{R+1})^{-\frac{2}{3}}-1\big\}}
\end{equation}
where $R=-12T_3^2/T_2^3$.
It is hard to believe at first, but the above superpotential is equivalent to the superpotential for $k=3$ in \ref{Eq:Sp_superpotential}.
One is described in terms of gauge invariants, while the other is described in the configuration of the quark fields that satisfy the D-flat conditions in the Wess-Zumino gauge.
The equivalence of the two is directly shown using Cardano's formula.

\subsection{Truly Confinement}

Here, we explain that this theory is a $t$-confining theory.
Furthermore, we determine that, as in the theories we previously studied, condensation of a hybrid ``magnetic'' operator and the breaking of discrete symmetry occur.

\paragraph{Unscreened and Index cConditions}

The first two conditions can be easily verified.
It is known that the center of the symplectic group is $\bZ_2$, and since the matter fields in the antisymmetric representation have charge $2$, screening does not occur.
The index of the adjoint representation is $I(\adj)=2(k+1)$, while the index of the antisymmetric representation is $I(A^{ij})=2(k-1)$ (see appendix \ref{sec:data_of_Lie_algebra}).
Therefore, it is clear that the index condition $I(A^{ij})<I(\adj)+2$ is satisfied.

\paragraph{Anomaly Matching Conditions}

When the signs of all branches in the superpotential are the same, the superpotential vanishes, and the moduli space includes the origin.  
That is, for any $k$, a branch with $W_\mathrm{eff}=0$ exists, and since the global symmetry is not broken at the origin, the anomaly matching conditions including the discrete symmetries can be verified.

\begin{table}[H]
    \centering
    \renewcommand{\arraystretch}{2}
    \scriptsize
    \begin{tabular}{p{2.4cm}|p{4.2cm}|p{3.5cm}|p{2.5cm}}
        anomaly & $\mathrm{UV}$ & $\mathrm{IR}$ & $\mathrm{UV}-\mathrm{IR}$ \\
        \hline
        $\tr U(1)_R(\mathrm{gravity})^2$ & $1\times k(2k+1)+\left(-\dfrac{2}{k-1}-1\right)\times\{k(2k-1)-1\}$ & $\sum_{i=2}^k\left(-\dfrac{2i}{k-1}-1\right)$ & $0$ \\
        $\tr U(1)_R^3$ & $1^3\times k(2k+1)+\left(-\dfrac{2}{k-1}-1\right)^3\times\{k(2k-1)-1\}$ & $\sum _{i=2}^k\left(-\dfrac{2i}{k-1}-1\right)^3$ & $0$ \\
        $\tr\bZ_{2k-2}(\mathrm{gravity})^2$ & $2\times\left[1\times\{k(2k-1)-1\}\right]$ & $2\times\left[\sum_{i=2}^ki\right]$ & $3k(k-1)$ \\
        $\tr\bZ_{2k-2}U(1)_R^2$ & $1\{-2-(k-1)\}^2\times\{k(2k-1)-1\}$ & $\sum_{i=2}^ki\{-2i-(k-1)\}^2$ & $-\dfrac{5}{6}k(k+1)(k-1)^2$ \\
        $\tr\bZ_{2k-2}^2U(1)_R$ & $1^2\{-2-(k-1)\}\times\{k(2k-1)-1\}$ & $\sum_{i=2}^ki^2\{-2i-(k-1)\}$ & $\dfrac{5}{6}k(k+1)(k-1)^2$ \\
        $\tr\bZ_{2k-2}^3$ & $1^3\times\{k(2k-1)-1\}$ & $\sum_{i=2}^ki^3$ & $-\dfrac{1}{4}k(k+4)(k-1)^2$ \\
    \end{tabular}
    \caption{anomaly matching conditions for $Sp(k)+A^{ij}$}
\end{table}
\noindent
Note that in the fourth and fifth rows of the table, the $R$-charges are normalized to be integers.
When $k$ is odd, the $R$-charges can be further normalized by a factor of one-half.

The anomaly matching conditions for these discrete symmetries cannot be satisfied assuming the discrete symmetry is $\bZ_{2k-2}$, so it is expected that it spontaneously breaks to $\bZ_{k-1}$.
Below, we explain why the anomaly of the discrete symmetry should match with the subgroup $\bZ_{k-1}$.

\paragraph{Dynamics of the Confinement Phase}

Here, we analyze the dynamics in the branch where $W_\mathrm{eff}=0$.
The discussion of anomaly matching in the previous section suggests that confinement leads to an effective theory at low energy, which is properly described by gauge invariant quantities.
Additionally, by decoupling flavors from the $s$-confining theory, it can be shown that the breaking of discrete symmetries occurs when a quark-gluon mixture acquires a vacuum expectation value in the confinement phase.

First, as in the previous section, consider the extension to an $s$-confining theory by adding quarks in the fundamental representation, $Q^i$ \cite{Csaki:1996sm,Csaki:1996zb}: $Sp(k)+A^{ij}+6Q^i$.
In $s$-confining theories, the form of the superpotential is well known.
Defining the composite operator $M_i=QA^iQ$, we can include a term in the superpotential for the $s$-confining theory
\begin{equation}
    W_\mathrm{eff}\supset\dfrac{1}{\Lambda^{2k+1}}M_0M_{k-1}^2.
\end{equation}
Introducing a mass term $mM_0$ leads to the vacuum expectation value of $M_{k-1}$.
The fact that $M_{k-1}$ acquires an expectation value triggers the condensation of a ``magnetic'' operator.
Let the basis for the representation space of the gauge group's antisymmetric representation be $E^i$, where the completeness relation $\Sigma_iE^iE^i=1$ holds.
The generalized Konishi anomaly for the infinitesimal transformation of $Q$ is given by
\begin{equation}
    \bar{D}^2(Q^\dagger E^iQ)=mQE^iQ+\dfrac{1}{16\pi^2}\tr(E^i\cW^\alpha\cW_\alpha).
\end{equation}
It is clear that the vacuum expectation value of the left-hand side is trivial.
The vacuum expectation value of $M_{k-1}$ leads to the condensation of a ``magnetic'' operator:
\begin{subequations}
\begin{align}
    M_{k-1}&=QA^{k-1}Q \\
    &=QE^iQ\tr(E^iA^{k-1}) \\
    &\propto\tr(E^i\cW^\alpha\cW_\alpha)\tr(E^iA^{k-1}) \\
    &=\tr(A^{k-1}\cW^\alpha\cW_\alpha)
\end{align}
\end{subequations}
The $\tr(A^{k-1}\cW^\alpha\cW_\alpha)$ has the $\bZ_{2k-2}$-charge $k-1$, and it inducesthe breaking of the discrete symmetry:
\begin{equation}
    \bZ_{2k-2}\to\bZ_{k-1}.
\end{equation}
Previously, we confirmed that the anomaly matching condition holds under $\bZ_{k-1}$, and we have now derived a direct justification for it.
In the confined phase of the $t$-confining theory, the condensation of a ``magnetic'' operator and the breaking of discrete symmetry indeed occur simultaneously.
At this point, one may wonder whether $t$-confinement is triggered by the condensation of actual magnetic degrees of freedom. Next, we partially clarify this point.

\paragraph{Condensation of a Magnetic Object}

It is believed that the confined phase arises from the condensation of magnetic objects, and in certain cases, this can be demonstrated in the $t$-confining theory.
One example is the work of Seiberg and Witten \cite{Seiberg:1994rs}, which was the first to demonstrate that monopole condensation leads to confinement.
Here, we demonstrate that the dual Meissner effect is indeed realized in the case of $k=3$.

Here is the strategy.
We start with $Sp(3)+2A^{ij}$ and decouple $A_2$ by introducing a mass term.
In this context, we label the two antisymmetric tensor fields as $A_1$ and $A_2$.
The next step is to explore the D-flat direction of $A_1$, which leads to the breaking of $Sp(3)\to SU(2)^3$.
The corresponding curve is well-known from \cite{Csaki:1997zg}.
In addition, we use the effective superpotential for monopoles, combined with the mass term for $A_2$.
This procedure allows us to identify monopole condensation, which, as shown in \cite{Intriligator:1994sm}, can be associated with the gaugino condensation.

Generally, along the D-flat direction of $A_1=\text{diag}(a_1, a_2, a_3)\otimes i\sigma_2$, gauge symmetry is broken as $Sp(3)\to SU(2)^3$.
The matching of the dynamical scales is given as follows:
\begin{equation}
    \Lambda_1^{4} = \frac{\Lambda^8}{(a_2 - a_1)^2 (a_3 - a_1)^2}, \,
    \Lambda_2^{4} = \frac{\Lambda^8}{(a_1 - a_2)^2 (a_3 - a_2)^2}, \,
    \Lambda_3^{4} = \frac{\Lambda^8}{(a_1 - a_3)^2 (a_2 - a_3)^2},
\end{equation}
where $\Lambda$ is the scale of $Sp(3)$ and $\Lambda_{1,2,3}$ are the scale of $SU(2)$s.
The curve for $SU(2)^3$ is
\begin{equation}
    y^2 = \left[ x^2 - \left( \Lambda_1^4 M_2 + \Lambda_2^4 M_3 + \Lambda_3^4 M_1 + T^2 - M_1 M_2 M_3 \right) \right]^2 - 4 \Lambda_1^4 \Lambda_2^4 \Lambda_3^4,
\end{equation}
where $M_i=Q_iQ_i$ is made of $A_2$ where $Q_i$ is the bifundamental under $SU(2)_i$ and $SU(2)_{i+1}$.
To obtain the superpotential, it is necessary to calculate the discriminant $\Delta$ of the right-hand side of the above equation.
The result of factoring $\Delta$ is as follows:
\begin{equation}
    \Delta=4096\Lambda_1^4\Lambda_2^4\Lambda_3^4\Delta_+\Delta_-
\end{equation}
where $\Delta_\pm=\Lambda_1^4 M_2 + \Lambda_2^4 M_3 + \Lambda_3^4 M_1 + T^2 - M_1 M_2 M_3 \pm2\Lambda_1^2\Lambda_2^2\Lambda_3^2$.
Therefore, the effective superpotential is given by the following expression
\begin{equation}
    W_\mathrm{eff}=\Delta_\pm q_\pm^2-m(M_1+M_2+M_3),
\end{equation}
where $q_\pm$ is treated as a new magnetic degree of freedom and $m$ is the mass of $A_2$.
All that remains is to integrate out $A_2$.
After performing this integration, the superpotential \ref{Eq:Sp_superpotential} is indeed reproduced.
This result demonstrates that the $t$-confining theory is triggered by the condensation of magnetic degrees of freedom.

\section{$SU(2k)$ with $A^{ij}$ and $\tilde{A}_{ij}$}
\label{sec:SU(2k)+A+A}

Here, we analyze the final model of the $t$-confining theory.
We have an antisymmetric tensor field $A^{ij}$ and its complex conjugate $\tilde{A}_{ij}$ in the gauge group $SU(2k)$.
This model is similar to the one considered in the previous section.
Therefore, much of the analysis can be carried out using the same methods.
First of all, the charges of the fields contained in the theory are as follows:
\begin{table}[H]
    \centering
    \myspace{1.5}
    \begin{tabular}{ScScScScSc}
        \hline
        symmetry & $SU(2k)$ & $U(1)_R$ & $U(1)_V$ & $\bZ_{4k-4}$ \\
        \hline
        $\cW^\alpha$\rule[-1mm]{0mm}{5mm} & $\adj$ & $1$ & $0$ & $0$ \\
        $A$ & $\ydiagram{1,1}$ & $-\dfrac{1}{k-1}$ & $1$ & $1$ \\
        $\tilde{A}$ & $\overline{\ydiagram{1,1}}$ & $-\dfrac{1}{k-1}$ & $-1$ & $1$ \\
        \hline
        $\tr(A\tilde{A})^i$ & $\one$ & $-\dfrac{2i}{k-1}$ & $0$ & $2i$ \\
        $\Pf A$ & $\one$ & $-\dfrac{k}{k-1}$ & $k$ & $k$ \\
        $\Pf\tilde{A}$ & $\one$ & $-\dfrac{k}{k-1}$ & $-k$ & $k$ \\
        \hline
    \end{tabular}
    \caption{$SU(2k)+A^{ij}+\tilde{A}_{ij}$}
\end{table}
The axial $U(1)$ symmetry of the antisymmetric matters is expected to break to $\bZ_{2I(A)}\simeq\bZ_{4k-4}$ due to instantons.
However, as shown through anomaly matching conditions for discrete symmetries and the subsequent discussion of the condensation of a ``magnetic'' operator, it is actually broken to $\bZ_{2k-2}$.
The classical moduli space is spanned by the following eigenvalues:
\begin{subequations}
\label{Eq:SU(2n)_moduli}
\begin{align}
    A&=
    \begin{pmatrix}
        a_1 && \\
        & \ddots & \\
        && a_n
    \end{pmatrix}
    \otimes\sigma_2 \\
    \tilde{A}&=
    \begin{pmatrix}
        \tilde{a}_1 && \\
        & \ddots & \\
        && \tilde{a}_n
    \end{pmatrix}
    \otimes\sigma_2
\end{align}
\end{subequations}
where $|a_i|^2-|\tilde{a}_i|^2=c$.
This leads to the breaking of the gauge group $SU(2k)$ to $SU(2)^k$.
The classical moduli space, expressed in terms of gauge invariant operators, is given by
\begin{subequations}
\begin{align}
    & \tr(A\tilde{A})^i \\
    & \Pf A \\
    & \Pf\tilde{A}
\end{align}
\end{subequations}
Here, $i=1,\cdots,k-1$.
There are no relations among the gauge invariant polynomials.
The term $\tr(A\tilde{A})^k$ is not an independent invariant and is therefore not included.
Through the super Higgs mechanism, $(4k^2-1)-3k=4k^2-3k-1$ massless chiral superfields are eaten by massless vector superfields.
Thus, the dimension of the classical moduli space is $\{2k(2k-1)\}-(4k^2-3k-1)=k+1$, which matches the dimensions found by the D-flatness conditions and gauge invariant polynomials.

\subsection{Superpotential}

As in the previous section, here we determine the dynamically generated superpotential and verify its consistency.
The non-perturbatively generated superpotential can be determined from symmetry and holomorphy:
\begin{equation}
\label{Eq:SU_superpotential}
    W_\mathrm{eff}=\sum_i\epsilon_i\left(\dfrac{\Lambda^{4k+2}}{\prod_{j\neq i}(a_i\tilde{a}_i-a_j\tilde{a}_j)^2}\right)^{\frac{1}{2}}
    =\sum_i\epsilon_i\dfrac{\Lambda^{2k+1}}{\prod_{j\neq i}(a_i\tilde{a}_i-a_j\tilde{a}_j)}
\end{equation}
where $\epsilon_i=\pm1$.
Note that the coefficient of the beta function is given by $4k+2$.
From the action of the Weyl group, it is understood that interchanging $a_i\tilde{a}_i$ corresponds to a gauge equivalent transformation, and the transformation that swaps the signs of $\epsilon_i$ due to the $R$-symmetry is also equivalent.
However, there still exist branches that can be distinguished despite these equivalences.
In particular, when all the signs of $\epsilon_i$ are the same, a branch with $W_\mathrm{eff}=0$ emerges.
The detailed computation is presented in appendix \ref{sec:W=0}.
The consistency of this superpotential can be verified through multiple methods.

\paragraph{Super Higgs Mechanism}

By performing the matching of the dynamical scales through the super Higgs mechanism, the superpotential can be derived.
In fact, by calculating the fermion masses from the Kähler potential $\tr\{A(e^V)^\top A^\dagger e^V\}+\tr\{\tilde{A}e^{-V}\tilde{A}^\dagger(e^{-V})^\top\}$, the holomorphic mass of the vector multiplet that acquires mass via the super Higgs mechanism can be computed.
When the gauge symmetry is spontaneously broken, the square of the holomorphic mass of the multiplet, which is in the fundamental representation of $SU(2)_i$ and the anti-fundamental representation of $SU(2)_j$, is given by $m_{i,j}^2=a_i\tilde{a}_i-a_j\tilde{a}_j$.
Thus, the matching of the dynamical scale for each $SU(2)_i$ factor is given by
\begin{equation}
    \dfrac{\Lambda^{4k+2}}{\prod_{j\neq i}m_{i,j}^4}=\dfrac{\Lambda^{4k+2}}{\prod_{j\neq i}(a_i\tilde{a}_i-a_j\tilde{a}_j)^2}=\Lambda_i^6.
\end{equation}
This result supports \ref{Eq:SU_superpotential}.
In addition, this superpotential reflects the gaugino condensation of each $SU(2)_i$ factor:
\begin{equation}
    W_\mathrm{eff}=\sum_i\pm(\Lambda_i^6)^{\frac{1}{2}}.
\end{equation}
The choice of signs for each factor leads to multiple branches.

\paragraph{Decoupling Colors}

Here, we consider the connection to theories with small $k$.
It is expected that the theory with $k-1$ is reproduced in the limit where one of the eigenvalues in \ref{Eq:SU(2n)_moduli} is large.
We assign one large vacuum expectation value to the $a_k=\tilde{a}_k$ and check the consistency with the case of the $SU(2(k-1))$ group.
The dynamical scale matching condition is given by $\Lambda^b=(\Lambda_L)^{b_L}{(a_k\tilde{a}_k)}^2$, where $\Lambda_L$ is the dynamical scale in the low-energy theory and $b_L$ is its beta function's coefficient.
\begin{align}
    W_\mathrm{eff}
    &=\sum_i\epsilon_i\dfrac{\Lambda^{\frac{b}{2}}}{\prod_{j\neq i}(a_i\tilde{a}_i-a_j\tilde{a}_j)}\\
    &=\sum_{i\neq k}\epsilon_i\dfrac{\Lambda^{\frac{b_L}{2}}{a_k\tilde{a}_k}}{\{\prod_{j\neq i,k}(a_i\tilde{a}_i-a_j\tilde{a}_j)\}(a_i\tilde{a}_i-a_k\tilde{a}_k)}+\epsilon_k\dfrac{\Lambda^{\frac{b_L}{2}}{a_k\tilde{a}_k}}{\prod_{j\neq k}(a_k\tilde{a}_k-a_j\tilde{a}_j)}\\
    &\to\sum_{i\neq k}{\epsilon_i}^{'}\dfrac{\Lambda^{\frac{b_L}{2}}}{\prod_{j\neq i,k}(a_i\tilde{a}_i-a_j\tilde{a}_j)},
\end{align}
where we considered the limit as $a_k$ and $\tilde{a}_k$ tend to infinity in the third line. We also redefined the signs ${\epsilon_i}^{'}$ as $-\epsilon_i$. 
This result corresponds to the case of $SU(2(k-1))$ group.
Therefore, it can be seen that there is consistency with decoupling colors and the change in $k$.

\paragraph{Decoupling Flavors}

We examine the behavior of the theory when mass is given to the quarks for small $k$, and verify that the number of vacua matches the $2k$ vacua in pure Yang-Mills theory.
As in the case of $Sp(k)$, we consider only the solutions that reproduce the gaugino condensation in pure Yang-Mills theory.
The superpotential, including mass terms, is given by
\begin{equation}
    W_\mathrm{eff}=\Lambda^{2k+1}\sum_i\dfrac{\epsilon_i}{\prod_{j\neq i}(a_i\tilde{a}_i-a_j\tilde{a}_j)}+\dfrac{1}{2}m\sum_ia_i\tilde{a}_i
\end{equation}
The vacuum solutions are obtained by solving the system of equations:
\begin{equation}
\label{Eq:sol}
    \dfrac{\partial W_\mathrm{eff}}{\partial a_i}=\dfrac{\partial W_\mathrm{eff}}{\partial\tilde{a}_i}=0\,,\quad(i=1,\cdots,k)
\end{equation}
The matching of the dynamical scale is given by $\Lambda_\mathrm{YM}^{6k}=m^{2k-2}\Lambda^{4k+2}$.
Thus, the gaugino condensation in pure Yang-Mills theory, when including mass $m$, is given by $W_\mathrm{eff}=(2k)(m^{2k-2}\Lambda^{4k+2})^{\frac{1}{2k}}$.
We need to reproduce this expression.
Vacuum solutions that do not satisfy this equation arise when one of the eigenvalues becomes degenerate, preventing the gauge group from breaking down to $SU(2)^k$, and are therefore excluded.
Additionally, vacua that are identified by the Weyl group of the gauge group are not distinguished.

Under the above strategy, consistency with Yang-Mills theory can be confirmed.
For example, for $k=2$, when mass term is added, the superpotential can be expressed in terms of mesons as
\begin{equation}
    W_\mathrm{eff}=(\epsilon_1+\epsilon_2)\left(\dfrac{\Lambda^{10}}{M^{11}M^{22}-(M^{12})^2}\right)^{\frac{1}{2}}+mM^{12}
\end{equation}
The solutions to the system of equations \ref{Eq:sol} exist when $\epsilon_1=\epsilon_2$, and are given by
\begin{equation}
    M^{11}=M^{22}=0\,,\quad M^{12}=(\omega_8)^i\sqrt{\dfrac{\Lambda^5}{m}}\,,\quad(i=1,3,5,7)
\end{equation}
This setup correctly reproduces the Witten index of pure Yang–Mills theory.
The mass terms are introduced because, in the limit $m\to0$, the vacuum would be pushed to infinity, and the Witten index would no longer be reproduced.
For the case of $SO(6)$, it suffices to add $m_1M^{11}+m_2M^{22}$ to the superpotential.
This result can also be derived without expressing the superpotential in terms of mesons, which are gauge-invariant quantities.
When the mass-deformed superpotential is expressed in terms of eigenvalues, the solutions to the system of equations excluding the identification under the action of the Weyl group are given by:
\begin{equation}
    a_1=\tilde{a}_1=0\,,\quad a_2\tilde{a}_2=(\omega_4)^i\sqrt{\dfrac{4\Lambda^5}{m}}\,,\quad(i=0,1,2,3)
\end{equation}
This again correctly reproduces the Witten index of pure Yang-Mills theory.
Of course, the result of the calculation remains unchanged whether the moduli space is described in terms of the eigenvalues of \ref{Eq:SU(2n)_moduli} or in terms of gauge invariant quantities.
The case for $k \geq 3$ can be analyzed in a similar manner.
The results obtained agree with those from pure Yang–Mills theory.

\paragraph{Consistency with $SO(6)+2V^i$}

For $k=2$, the theory $SU(4)+A^{ij}+\tilde{A}_{ij}$ is locally equivalent to the supersymmetric gauge theory $SO(6)+2V^i$, so we can check the consistency by verifying whether the results obtained in \cite{Intriligator:1995id} match.
Note that $\mathfrak{su(4)\simeq\mathfrak{so}(6)}$, and the antisymmetric representation of $\mathfrak{su}(4)$ is equivalent to its complex conjugate and to the vector representation of $\mathfrak{so}(6)$.
We make use of the results from sec. \ref{sec:SO(N)+(N-4)V} once again.
The superpotential for $SO(6)+2V^i$ is given by
\begin{equation}
    W_\mathrm{eff}\propto\left(\dfrac{\Lambda^{10}}{\det M^{\{ij\}}}\right)^\frac{1}{2}.
\end{equation}
Here, if we express the quark fields in the vector representation as $V^{\alpha i}$, we have
\begin{equation}
    \det M^{\{ij\}}=\det\dfrac{1}{2}\delta_{\alpha\beta}V^{\alpha i}V^{\beta j}.
\end{equation}
In terms of the eigenvalues of \ref{Eq:SU(2n)_moduli}, we obtain
\begin{subequations}
\begin{align}
    \det M^{\{ij\}}&=-\det
    \begin{vmatrix}
        2a_1a_2 & a_1\tilde{a}_1+a_2\tilde{a}_2 \\
        a_1\tilde{a}_1+a_2\tilde{a}_2 & 2\tilde{a}_1\tilde{a}_2
    \end{vmatrix} \\
    &=(a_1\tilde{a}_1-a_2\tilde{a}_2)^2.
\end{align}
\end{subequations}
Thus, the superpotential becomes
\begin{equation}
    W_\mathrm{eff}\propto\left(\dfrac{\Lambda^{10}}{(a_1\tilde{a}_1-a_2\tilde{a}_2)^2}\right)^{\frac{1}{2}},
\end{equation}
which matches the superpotential for $SU(4)+A^{ij}+\tilde{A}_{ij}$ in \ref{Eq:SU_superpotential}.

\paragraph{Consistency with $Sp(k)+A^{ij}$}

By considering a specific direction in the moduli space, we should be able to recover the superpotential of theory $Sp(k)+A^{ij}$ discussed in the previous section.
This is because, assuming that all $\tilde{a}_i$ take the same value, the gauge group breaks down to $Sp(k)$.
One quark remains in the antisymmetric representation, while quarks in the trivial representation do not contribute to the dynamics and are ignored. 
This way, we can check if $Sp(k)+A^{ij}$ can be reproduced by the super Higgs mechanism.

We place all $\tilde{a}_i$ as $\tilde{a}$ and integrate them.
The matching of the dynamical scale is given by
\begin{equation}
    \Lambda_{SU(2k)}^{4k+2}=\Lambda_{Sp(k)}^{2k+4}\tilde{a}^{2k-2}
\end{equation}
 where we use the one loop coefficients of the beta function: $b_{SU(2k)}=4k+2$ and $b_{Sp(k)}=2k+4$.
The superpotential is then given by
\begin{subequations}
\begin{align}
W_\mathrm{eff}&=\sum_i\epsilon_i\left(\dfrac{\Lambda_{SU(2k)}^{4k+2}}{\prod_{j\neq i}(a_i\tilde{a}_i-a_j\tilde{a}_j)^2}\right)^{\frac{1}{2}}
=\sum_i\epsilon_i\left(\dfrac{\Lambda_{Sp(k)}^{2k+4}\tilde{a}^{2k-2}}{\prod_{j\neq i}(a_i\tilde{a}_i-a_j\tilde{a}_j)^2}\right)^{\frac{1}{2}}\notag\\
&=\sum_i\epsilon_i\left(\dfrac{\Lambda_{Sp(k)}^{2k+4}\tilde{a}^{2k-2}}{\prod_{j\neq i}\{(a_i-a_j)\tilde{a}\}^2}\right)^{\frac{1}{2}}
    =\sum_i\epsilon_i\left(\dfrac{\Lambda_{Sp(k)}^{2k+4}}{\prod_{j\neq i}(a_i-a_j)^2}\right)^{\frac{1}{2}}.
\end{align}
\end{subequations}
This reproduces the previous result in \ref{Eq:Sp_superpotential}.

\subsection{Truly Confinement}

Here, we explain that this theory is a $t$-confining theory.
There are three conditions, so we examine them one by one.

\paragraph{Unscreened and Index Conditions}

We now check whether this theory satisfies the conditions for a $t$-confining theory.
The center of the Lie group $SU(2k)$ is $\mathbb{Z}_{2k}$, and the matter fields in the antisymmetric representation carry center charge $2$.
Clearly, the greatest common divisor of $2k$ and $2$ is $2$, indicating the presence of a nontrivial center symmetry.
Furthermore, the Wilson line in the fundamental representation exhibits an area law, thus satisfying the first condition.
The index condition can be verified through a straightforward calculation.
The table in the appendix \ref{sec:data_of_Lie_algebra} provides the relevant details.
The Dynkin index is $2(2k)$ for the adjoint representation and $2k-2$ for the antisymmetric representation.
The index condition corresponds to $2\times(2k-2)<2(2k)+2$, and this equation is indeed satisfied.
Therefore, we see that both unscreened and index conditions are satisfied.

\paragraph{Anomaly Matching Conditions}

When all the signs of the branches in the superpotential are the same, the superpotential vanishes, and the origin is included in the moduli space.
We expect that for any $k$, the anomaly matching conditions including those for the discrete symmetry can be verified.
\begin{table}[H]
    \centering
    \renewcommand{\arraystretch}{2}
    \scriptsize
    \begin{tabular}{p{2.4cm}|p{4.5cm}|p{3.6cm}|p{2.5cm}}
        anomaly & $\mathrm{UV}$ & $\mathrm{IR}$ & $\mathrm{UV}-\mathrm{IR}$ \\
        \hline
        $\tr U(1)_R(\mathrm{gravity})^2$ & $1\times(4k^2-1)+\left(-\dfrac{1}{k-1}-1\right)\times2k(2k-1)$ & $\sum_{i=1}^{k-1}\left(-\dfrac{2i}{k-1}-1\right)+2\left(-\dfrac{k}{k-1}-1\right)$ & $0$ \\
        $\tr U(1)_R^3$ & $1^3\times(4k^2-1)+\left(-\dfrac{1}{k-1}-1\right)^3\times2k(2k-1)$ & $\sum_{i=1}^{k-1}\left(-\dfrac{2i}{k-1}-1\right)^3+2\left(-\dfrac{k}{k-1}-1\right)^3$ & $0$ \\
        $\tr U(1)_RU(1)_V^2$ & $\left(-\dfrac{1}{k-1}-1\right)1^2\times k(2k-1)+\left(-\dfrac{1}{k-1}-1\right)(-1)^2\times k(2k-1)$ & $\left(-\dfrac{k}{k-1}-1\right)k^2+\left(-\dfrac{k}{k-1}-1\right)(-k)^2$ & $0$ \\
        $\tr\bZ_{4k-4}\mathrm{(gravity)}^2$ & $1\times2k(2k-1)$ & $\sum_{i=1}^{k-1}2i+k\times2$ & $3k(k-1)$ \\
        $\tr\bZ_{4k-4}U(1)_V^2$ & $1\cdot1^2\times k(2k-1)+1\cdot(-1)^2\times k(2k-1)$ & $k\cdot k^2+k(-k)^2$ & $-2k(k-1)^2$ \\
        $\tr\bZ_{4k-4}U(1)_R^2$ & $1\{-1-(k-1)\}^2\times2k(2k-1)$ & $\sum_{i=1}^{k-1}2i\{-2i-(k-1)\}^2+k\{-k-(k-1)\}^2\times2$ & $-\dfrac{1}{3}k(k-1)^2(5k-1)$ \\
        $\tr\bZ_{4k-4}^2U(1)_R$ & $1^2\{-1-(k-1)\}\times2k(2k-1)$ & $\sum_{i=1}^{k-1}(2i)^2\{-2i-(k-1)\}+k^2\{-k-(k-1)\}\times2$ & $\dfrac{2}{3}k(k-1)^2(5k-1)$ \\
        $\tr\bZ_{4k-4}^3$ & $1^3\times2k(2k-1)$ & $\sum_{i=1}^{k-1}(2i)^3+k^3\times2$ & $-2k(k-1)^2(k+1)$ \\
    \end{tabular}
    \caption{Anomaly matching conditions for $SU(2k)+A^{ij}+\tilde{A}_{ij}$}
\end{table}
\noindent
Only the nontrivial anomalies are listed in the table above.
The matching of the anomalies $\tr U(1)_V^3$, $\tr U(1)_V\mathrm{(gravity)}^2$, $\tr U(1)_R^2U(1)_V$, and $\tr\bZ_{4k-4}U(1)_RU(1)_V$ is obvious.
Since these anomaly matching conditions for the discrete symmetries cannot be satisfied when assuming the discrete symmetry is $\bZ_{4k-4}$, it is expected that the symmetry spontaneously breaks to $\bZ_{2k-2}$.
Below, we show that the discrete symmetry is broken, and consequently, demonstrate that the anomalies match completely.

\paragraph{Dynamics of the Confinement Phase}

The discussion of the anomaly matching conditions in the previous section suggests that the low-energy effective theory in the confined phase is appropriately described by gauge invariant quantities.
We analyze the dynamics in the confined phase by decoupling the flavors from the $s$-confined theory, we can show that the quark-gluon mixture acquires a vacuum expectation value, which leads to the breaking of the discrete symmetry.

First, consider the extension to the $s$-confining theory by adding quarks in the fundamental representation and its complex conjugate \cite{Csaki:1996sm,Csaki:1996zb}: $SU(2k)+A^{ij}+\tilde{A}_{ij}+3(Q^i+\tilde{Q}^i)$.
In $s$-confining theories, the superpotential is constrained in such a way that it leads to the equations of motion, and its form is well-known.
If we define the composite operator $M_i=\tilde{Q}(A\tilde{A})^iQ$, the $s$-confining theory contains a term in the superpotential:
\begin{equation}
    W_\mathrm{eff}\supset\dfrac{1}{\Lambda^{4k-1}}M_0M_{k-1}^2.
\end{equation}
Introducing a mass term $mM_0$, the field $M_{k-1}$ acquires a vacuum expectation value.
Let the basis of the representation space of the adjoint representation of the gauge group be $E^i$, and assume the completeness relation $\Sigma_iE^iE^i=1$ holds.
The generalized Konishi anomaly for $Q$ under an infinitesimal transformation given by $E^iQ$ is expressed as
\begin{equation}
    \bar{D}^2(Q^\dagger E^iQ)=m\tilde{Q}E^iQ+\dfrac{1}{16\pi^2}\tr(E^i\cW^\alpha\cW_\alpha).
\end{equation}
The left-hand side is trivial, thus the condensation of the ``magnetic'' operator is given by
\begin{subequations}
\begin{align}
    M_{k-1}&=\tilde{Q}(A\tilde{A})^{k-1}Q \\
    &=\tilde{Q}E^iQ\tr\{E^i(A\tilde{A})^{k-1}\} \\
    &\propto\tr(E^i\cW^\alpha\cW_\alpha)\tr\{E^i(A\tilde{A})^{k-1}\} \\
    &=\tr\{(A\tilde{A})^{k-1}\cW^\alpha\cW_\alpha\}.
\end{align}
\end{subequations}
The $\tr\{(A\tilde{A})^{k-1}\cW^\alpha\cW_\alpha\}$ has the $\bZ_{4k-4}$-charge $2k-2$, and when it acquires a vacuum expectation value, the discrete symmetry $\bZ_{4k-4}$ breaks:
\begin{equation}
    \bZ_{4k-4}\to\bZ_{2k-2}.
\end{equation}
This result shows that the anomaly matching conditions of the discrete symmetry are correct.
Furthermore, this theory also shows that the condensation of the ``magnetic'' operator and the breaking of the discrete symmetry occur simultaneously in the confinement phase.
We have observed properties that are common to many $t$-confining models.
These are summarized in the next discussion.

\section{$SO(12)$ with two spinors}
\label{sec:SO(12)+2s}

Here, we analyze the final model of the $t$-confining theory.
We have two spinor representations $S$ of $SO(12)$. The center group of $SO(12)$ is $\bZ_2 \times \bZ_2$, and the spinor representation is odd under one $\bZ_2$, but even under the other $\bZ_2$. Therefore it cannot screen the entire center group and the theory is expected to be confining. 

One can consider the case with only one spinor. However in this case, $SO(12)$ is broken to $SU(6)$ and the theory has a run-away behavior with no ground state. The spinor of $SO(12)$ can be decomposed under $U(6)$ as $32 = 1_{3}+15_{1}+15^*_{-1}+1_{-3}$. Going along the direction of the singlets with vanishing $U(1)$ D-term, $SO(12)$ breaks to $SU(6)$, where $66-35=31$ fields are eaten out of 32. There is only one modulus $S^4$, and the superpotential is
\begin{align}
    W = \left(\frac{\Lambda^{26}}{(S^4)^2}\right)^{1/6}.
\end{align}
It clearly does not have a ground state.

On the other hand, with two spinors $S_i$ with $i=1,2$, we expect to find a moduli space of vacua. This is because under the breaking pattern above, the second $S$ is $15+15^*$ under the unbroken $SU(6)$ gauge group, which is one among the series of $SU(2k)+A^{ij}+\tilde{A}_{ij}$ with $k=3$. 

First of all, the charges of the fields contained in the theory are as follows:
\begin{table}[H]
    \centering
    \myspace{1.5}
    \begin{tabular}{ScScScScSc}
        \hline
        symmetry & $SO(12)$ & $SU(2)$ & $U(1)_R$ & $\bZ_{16}$ \\
        \hline
        $\cW^\alpha$\rule[-1mm]{0mm}{5mm} & $\adj$ & $1$ & $1$ & $0$ \\
        $S$ & 32 & 2 & $-\frac{1}{4}$ & $1$ \\
        \hline
        $S^2$ & $\one$ & $\one$ & $-\frac{1}{2}$ & $2$ \\
        $S^4$ & $\one$ & $\ydiagram{4}$ & $-1$ & $4$ \\
        $S^6$ & $\one$ & $\one$ & $-\frac{3}{2}$ & $6$ \\
        \hline
    \end{tabular}
    \caption{$SO(12)+2S$}
\end{table}

\subsection{Truly Confinement}

Here, we explain that this theory is a $t$-confining theory.
There are three conditions, so we examine them one by one.

\paragraph{Unscreened and Index Conditions}

We now check whether this theory satisfies the conditions for a $t$-confining theory.
The center of the Lie group $SO(12)$ is $\bZ_2 \times \bZ_2$, and the matter fields in the spinor representation carry center charge $(-,+)$. Therefore, they cannot screen all of the center charges and hence we expect the theory to be $t$-confining.
Note that we are strictly considering the center of $Spin(12)$, but do not too concerned about it.
Checking the index conditions is more easy.
The Dynkin index of the spinor representation is $8$, and that of the adjoint (antisymmetric) representation is $20$, where we refer to the table in the appendix \ref{sec:data_of_Lie_algebra}.
The index condition \ref{Eq:index_condition} is indeed satisfied when the number of spinors is either one or two.

\paragraph{Anomaly Matching Conditions}

The third condition is the anomaly matching conditions.
Here are the checks of the conditions.
Again we consider the conditions including the discrete symmetry.

\begin{table}[H]
    \centering
    \renewcommand{\arraystretch}{2}
    \scriptsize
    \begin{tabular}{p{2.5cm}|p{3.0cm}|p{4.6cm}|p{1.2cm}}
        anomaly & $\mathrm{UV}$ & $\mathrm{IR}$ & $\mathrm{UV}-\mathrm{IR}$ \\
        \hline
        $\tr U(1)_R(\mathrm{gravity})^2$ & $66\times 1 + 32 \times 2 \times (-\frac{5}{4})$ & $\left(-\frac{3}{2}\right)+5\times(-2)+\left (-\frac{5}{2}\right)$ & $0$ \\
        $\tr\bZ_{16}\mathrm{(gravity)}^2$ & $32 \times 2$ & $2+5\times 4+6$ & $36$ \\
       $\tr U(1)_R^3$ & $66 \cdot 1^3+32 \times 2 \left(-\frac{5}{4}\right)^3$ & $1\left( -\frac{3}{2}\right)^3 + 5 (-2)^3 + 1\left(-\frac{5}{2}\right)^3$ & $0$ \\
        $\tr\bZ_{16}U(1)_R^2$ & $32 \times 2(-5)^2$ & $2(-6)^2+5 \times 4(-8)^2+6(-10)^2$ & $-352$ \\
        $\tr\bZ_{16}^2U(1)_R$ & $32 \times 2\times(-5)$ & $2^2(-6)+5 \times 4^2(-8)+6^2(-10)$ & $704$ \\
        $\tr\bZ_{16}^3$ & $32 \times2$ & $2^3+5 \times 4^3+6^3$ & $-480$ \\
        $\tr U(1)_R SU(2)^2$ & $32 \times \frac{1}{2}\left(-\frac{5}{4}\right)$ & $10\times \left(-2\right)$ & $0$ \\
        $\tr\bZ_{16}SU(2)^2$ & $32 \times \frac{1}{2}\times1$ & $10\times 4$ & $-24$ \\
        $\tr\bZ_{16}SO(12)^2$ & $8\times2$ & $0$ & $16$ \\
    \end{tabular}
    \caption{Anomaly matching conditions for $SO(12)+2S$}
\end{table}
\noindent
Only the non-trivial anomalies are listed in the table above. $R$-charges are made integers by using $4R$ when considering the anomalies which include the discrete $\bZ_{16}$ symmetry. The anomalies of discrete $\bZ_{16}$ do not match. However, assuming that the ``magnetic'' $\tr(S^8\cW^\alpha\cW_\alpha)$ operator acquires an vacuum expectation value, $\bZ_{16}$ breaks to $\bZ_{8}$. Then all anomalies do match.
The condensation of such an operator is expected to be obtained by considering the upstream $s$-confining theory $SO(12)+2S+3V$ and decoupling the vector flavors.
Indeed, the $s$-confining superpotential of that theory contains $W_\mathrm{eff}\supset V^2(S^8V^2)^2/\Lambda^{19}$\cite{Csaki:1996zb}, and by adding the mass term $\tr(mV^2)$ for the vector, it is expected that $S^8V^2$ will condense inducing the condensation of $\tr(S^8\cW^\alpha\cW_\alpha)$, as in the previous case.

\section{Discussions}
\label{sec:discussions}

We have defined, classified, and analyzed $t$-confining theories.
In defining $t$-confining theories, we extracted key properties of the phenomenon of confinement.
First, in a strict sense, confinement requires that the potential between quarks becomes asymptotically linear.
In such cases, the vacuum expectation value of the Wilson loop operator follows an area law.
This necessitates that the center of the gauge group remains unscreened by matter fields, which we adopted as the first condition for a $t$-confining theory.
The second condition is somewhat empirical.
$s$-confining theories impose a Dynkin index condition on all matter fields included in the theory.
There are no known models in which confinement occurs when the total index exceeds that of an $s$-confining theory.
Therefore, requiring the index to be smaller than that of $s$-confining theories is reasonable.
The third condition is anomaly matching.
This is an obvious requirement when attempting to demonstrate confinement.
We also required the matching of anomalies for discrete symmetries.
This is sensitive to the presence of condensates and, in a certain sense, reflects the dual Meissner effect, where the condensation of ``magnetic'' objects leads to confinement.
In fact, in most cases, we could demonstrate the condensation of ``magnetic'' objects. Only for $SO(k)$ theory with $k-3$ vectors, a magnetic object could be found by adding an additional flavor. It leads to a Coulomb branch where massless monopoles exist at a singularity, and decoupling the extra flavor leads to the condensation of monopoles.

Next, we classified $t$-confining theories under the assumption that there is no tree-level superpotential and that the gauge group is simple.
When matter fields in the fundamental or spinor representations are present, they screen the center, so only larger representations should be included in the theory.\footnote{Except for $SO(4k)$ gauge theories}
However, this inevitably increases the Dynkin index of the matter fields, creating a tension with the index condition.
This fact led to the exclusion of many candidate theories, and as a result, only seven theories were classified as $t$-confining.

Among the seven theories classified as $t$-confining, we briefly reviewed those that have been studied in previous work.
There are three previously unexplored theories. Regarding two of the three theories, we derived the superpotential, checked its consistency, verified the anomaly matching conditions including those for discrete symmetries, and analyzed the condensation of ``magnetic'' operators as well as the breaking of discrete symmetries. As for the remaining one, we verified the anomaly matching conditions including those for discrete symmetries and the existence of the ``magnetic'' operator.

Through the analysis of the seven $t$-confining theories, several characteristic features have become clear.
In the theories classified as (A) through (G) in sec. \ref{sec:classification}, we have summarized the dynamics in the confined phases in the table below:
\begin{table}[H]
    \centering
    \myspace{2}
    \begin{tabular}{SlSlSrScSlSl}
        \hline
         theory & ``magnetic'' operator & \multicolumn{3}{Sc}{discrete symmetry} & reference \\
        \hline
        $G$ & $\tr(\cW^\alpha\cW_\alpha)$ & $\bZ_{2h}$ & $\to$ & $\bZ_2$ & \cite{Veneziano:1982ah,Witten:1982df} \\
        $SO(k)+(k-4)\,\ydiagram{1}$ & $\tr(V^{k-4}\cW^\alpha\cW_\alpha)$ & $\bZ_{2k-8}$ & $\to$ & $\bZ_{k-4}$ & \cite{Intriligator:1995id} and \cite{Csaki:1997aw} \\
        $SO(k)+(k-3)\,\ydiagram{1}$ & & & & & \cite{Intriligator:1995id} and \cite{Csaki:1997aw} \\
        $SU(6)+\ydiagram{1,1,1}$ & $\tr(A^2\cW^\alpha\cW_\alpha)$ & $\bZ_6$ & $\to$ & $\bZ_2$ & \cite{Henning:2021ctv} and \cite{Csaki:1997aw} \\
        $Sp(k)+\ydiagram{1,1}$ & $\tr(A^{k-1}\cW^\alpha\cW_\alpha)$ & $\bZ_{2k-2}$ & $\to$ & $\bZ_{k-1}$ & sec. \ref{sec:Sp(2k)+A} \\
        $SU(2k)+\ydiagram{1,1}+\overline{\ydiagram{1,1}}$ & $\tr\{(A\tilde{A})^{k-1}\cW^\alpha\cW_\alpha\}$ & $\bZ_{4k-4}$ & $\to$ & $\bZ_{2k-2}$ & sec. \ref{sec:SU(2k)+A+A} \\
        $SO(12)+2\,S$ & $\tr(S^8\cW^\alpha\cW_\alpha)$ & $\bZ_{16}$ & $\to$ & $\bZ_{8}$ & sec. \ref{sec:SO(12)+2s} \\
        \hline
    \end{tabular}
    \caption{Summary of dynamics in the confinement phase of $t$-confining theories}
\end{table}
First, in $t$-confining theories except for (A) and (G), there exists a branch with $W = 0$.
As noted earlier, it vanishes only effectively for $SO(k) + (k-3)\,\ydiagram{1}$ (see sec. \ref{sec:SO(N)+(N-3)V}).
Notably, all of these theories are multibranch theories, and it is remarkable that each of them realizes a $W = 0$ branch through a choice of signs in the superpotential.
Moreover, the condensation of ``magnetic'' operators and the associated spontaneous breaking of discrete symmetries are observed in most of $t$-confining phases, while in the case (C), there is no condensation and no symmetry breaking.
However, theory (C) lies on a branch that arises from the condensation of real magnetic monopoles in an upstream theory with an extra flavor.

Thus far, we have studied all aspects of $t$-confining theories.
It is interesting to note that they exhibit several common features.
In this work, we focused on $\mathcal{N} = 1$ supersymmetric gauge theories, but in the future, we expect to explore the phase structure of non-supersymmetric gauge theories as well employing anomaly-mediated supersymmetry breaking \cite{Randall:1998uk,Giudice:1998xp} following recent applications to non-perturbative dynamics \cite{Murayama:2021xfj,Csaki:2021xhi,Csaki:2021aqv,Csaki:2021jax,Csaki:2021xuc,Kondo:2021osz,Kondo:2022lvu,Leedom:2025mcg,Goh:2025oes}.

\appendix

\section{Details of Calculations}
\label{sec:calculation}

In this appendix, we summarize the basic calculations required in the main text.
First, we demonstrate that in the Secs.~\ref{sec:Sp(2k)+A} and \ref{sec:SU(2k)+A+A}, there exist branches where the superpotential vanishes.
Furthermore, we examine the action of the Weyl group on the matter fields in the antisymmetric representation.
Here, we limit ourselves to the above two fundamental points and omit other calculations.

\subsection{The Proof of $W=0$}
\label{sec:W=0}

In this section, we give the proof of the existence of a branch with non-trivially vanishing superpotential.
This branch is observed in theories with antisymmetric fields when the gauge group is $Sp(k)$ and $SU(2k)$ (see Secs.~\ref{sec:Sp(2k)+A} and \ref{sec:SU(2k)+A+A}).
Now we show two proofs, one is the direct computation and the other does not include complicated computation and is more powerful.

First, we show the existence of a $W=0$ branch with a direct computation.
The statement we proof is the following formula becomes zero:
\begin{equation}
\label{Eq:W=0}
    W=\sum_i\frac{1}{\prod_{j\neq i}(a_i-a_j)}=0,
\end{equation}
where the indexes $i,j$ take the value from $1$ to $k$.
For example, in the case of $k=3$, the superpotential becomes
\begin{equation}
    W^{(k=3)}=\frac{1}{(a_1-a_2)(a_1-a_3)}+\frac{1}{(a_2-a_1)(a_2-a_3)}+\frac{1}{(a_3-a_1)(a_3-a_2)},
\end{equation}
and by easy calculation, we can see above expression is equal to zero.
To prove this statement in general $k$, we first review Vandermonde matrix, its determinant and its inverse matrix.
The Vandermonde matrix is given by
\begin{equation}
    V:=
    \begin{pmatrix}
        1 & a_1 & a_1^2 & \cdots & a_1^{k-1} \\
        1 & a_2 & a_2^2 & \cdots & a_2^{k-1} \\
        1 & a_3 & a_3^2 & \cdots & a_3^{k-1} \\
        \vdots & \vdots & \vdots & \ddots & \vdots \\
        1 & a_k & a_k^2 & \cdots & a_k^{k-1}
    \end{pmatrix}.
\end{equation}
In addition, the determinant of the matrix, which is often called Vandermonde determinant, is given by
\begin{equation}
    \det V=\prod_{i<j}^k(a_i-a_j)
\end{equation}
Furthermore, these is a famous theorem that the inverse matrix of a squre matrix is given by its adjugate matrix divided by its determinant.
The $(i,j)$ element of the adjugate matrix is given by $(-1)^{i+j}M_{j,i}$ where $M_{i,j}$ is the determinant of the $(i,j)$-minor of the original matrix.
Note that in the case of Vandermonde matrix the $(i,k)$-minor where $1\leq i\leq k$ is given by the determinant of one dimension smaller Vandermonde matrix.
Thus, the adjugate matrix is given by
\begin{equation}
    \text{Adj}(V)=
    \begin{pmatrix}
        * & * & * & \cdots & * \\
        * & * & * & \cdots & * \\
        * & * & * & \cdots & * \\
        \vdots & \vdots & \vdots & \ddots & \vdots \\
        (-1)^{1+k}M_{1,k} & (-1)^{2+k}M_{2,k} & (-1)^{3+k}M_{3,k} & \cdots & (-1)^{k+k}M_{k,k}
    \end{pmatrix},
\end{equation}
The last row of the inverse matrix becomes
\begin{equation}
    V^{-1}_{k,i}=\frac{1}{\det V}\text{Adj}(V)_{k,i}=\frac{1}{\prod_{j\neq i}(a_i-a_j)}.
\end{equation}
So we get the relation
\begin{equation}
    W=\sum_iV^{-1}_{k,i}.
\end{equation}
Writing this relation in the matrix form, we obtain
\begin{equation}
    \begin{pmatrix}
        x_1 \\
        x_2 \\
        x_3 \\
        \vdots \\
        W
    \end{pmatrix}
    =V^{-1}
    \begin{pmatrix}
        1 \\
        1 \\
        1 \\
        \vdots \\
        1
    \end{pmatrix},
\end{equation}
where we use unknown variables $x_i$.
Multiplying both sides by the matrix V, the equation can be transformed as
\begin{equation}
    \begin{pmatrix}
        1 & a_1 & a_1^2 & \cdots & a_1^{k-1} \\
        1 & a_2 & a_2^2 & \cdots & a_2^{k-1} \\
        1 & a_3 & a_3^2 & \cdots & a_3^{k-1} \\
        \vdots & \vdots & \vdots & \ddots & \vdots \\
        1 & a_k & a_k^2 & \cdots & a_k^{k-1}
    \end{pmatrix}
    \begin{pmatrix}
        x_1 \\
        x_2 \\
        x_3 \\
        \vdots \\
        W
    \end{pmatrix}
    =
    \begin{pmatrix}
        1 \\
        1 \\
        1 \\
        \vdots \\
        1
    \end{pmatrix}.
\end{equation}
Since this equation holds for any $a_i$, it follows that $x_1=1$ and other variables including $W$ are zero.
Finally we conclude the equation \ref{Eq:W=0} is correct.

Second, we show $W=0$ by indirect proof.
When we assume $W\neq0$, we encounter a contradiction.
Assigning a mass dimension of one to $a_i$, then the mass dimension of the superpotential becomes $-(k-1)$, which is less than zero.
However, the superpotential has no pole and therefore the dimension should not be negative.
This is a contradiction, and we conclude $W=0$.
The proof ends here, but below, we explain why the superpotential does not have a pole.
It is sufficient to show that $a_i-a_j$ does not appear in the denominator of $W$ for any $i,j$.
Explicitly expressing only the terms where $a_i-a_j$ appears in the denominator, $W$ becomes as follows:
\begin{align}
    W&=\frac{1}{(a_i-a_j)\prod_{k\neq i,j}(a_i-a_k)}+\frac{1}{(a_j-a_i)\prod_{l\neq i,j}(a_j-a_l)}+\cdots\\
    &=\frac{\prod_{l\neq i,j}(a_j-a_l)-\prod_{k\neq i,j}(a_i-a_k)}{(a_i-a_j)\prod_{k\neq i,j}(a_i-a_k)\prod_{l\neq i,j}(a_j-a_l)}+\cdots
\end{align}
If we assume $a_i=a_j$, the numerator becomes zero, making factorization by $a_i-a_j$ possible.
Furthermore, since the denominator has only one factor of $a_i-a_j$, we conclude that the superpotential does not have a pole.

\subsection{The action of the Weyl group}
\label{sec:actions_of_the _Weyl_groups}

We aim to clarify how the eigenvalues of the classical moduli space which are given in \ref{Eq:moduli_Sp_asym} are transformed under the Weyl group transformations.
Here, we will compute both the Weyl group of the Lie algebra and the Weyl group of the Lie group.

First, let's determine the Weyl group of the $\mathfrak{sp}(k)$ algebra.
The simple roots of the $\mathfrak{sp}(k)$ algebra are given by:
\begin{equation}
    \begin{cases}
        \alpha^{(i)}=e_i-e_{i+1}\quad(1\leq i\leq k-1) \\
        \alpha^{(n)}=2e_n
    \end{cases}
\end{equation}
The root system of $\mathfrak{sp}(k)$ is spanned by the following:
\begin{equation}
    \begin{cases}
        \pm e_i\pm e_j\quad&(1\leq i\neq j\leq k) \\
        \pm2e_i&(1\leq i\leq k)
    \end{cases}
\end{equation}
The Weyl group is generated by reflections across hyperplanes perpendicular to the roots, so all elements can be enumerated. 

\noindent
(1) When the root is $\alpha=e_i+e_j$, any vector $v$ is transformed as:
\begin{align}
    v\to&v-2\dfrac{(v,\alpha)}{(\alpha,\alpha)}\alpha\\
    =&(v_1,\cdots,v_i,\cdots,v_j,\cdots,v_k)-2\dfrac{v_i+v_j}{2}(0,\cdots,1,\cdots,1,\cdots,0)\\
    =&(v_1,\cdots,-v_j,\cdots,-v_i,\cdots,v_k)
\end{align}
Thus, $v_i$ and $v_j$ are swapped and their signs are reversed.
The same transformation occurs when the root is $\alpha=-e_i-e_j$.

\noindent
(2) When the root is $\alpha=e_i-e_j$, any vector $v$ is transformed as:
\begin{align}
    v\to&v-2\dfrac{(v,\alpha)}{(\alpha,\alpha)}\alpha\\
    =&(v_1,\cdots,v_i,\cdots,v_j,\cdots,v_k)-2\dfrac{v_i-v_j}{2}(0,\cdots,1,\cdots,-1,\cdots,0)\\
    =&(v_1,\cdots,v_j,\cdots,v_i,\cdots,v_k)
\end{align}
Thus, $v_i$ and $v_j$ are swapped.
The same transformation occurs when the root is $\alpha=-e_i+e_j$.

\noindent
(3) When the root is $\alpha=2e_i$, any vector $v$ is transformed as:
\begin{align}
    v\to&v-2\dfrac{(v,\alpha)}{(\alpha,\alpha)}\alpha\\
    =&(v_1,\cdots,v_i,\cdots,v_k)-2\dfrac{2v_i}{4}(0,\cdots,2,\cdots,0)\\
    =&(v_1,\cdots,-v_j,\cdots,v_k)
\end{align}
Thus, the sign of $v_i$ is reversed. The same transformation occurs when the root is $\alpha=-2e_i$.
As a result, the Weyl group consists of permutations of the $k$ elements and sign reversals of each element.
As a group, it is isomorphic to $S_k\rtimes\bZ_2^k$.

Next, let's consider the group $Sp(k)$.
The generators of the maximal torus of $Sp(k)$ are given by:
\begin{equation}
    \begin{pmatrix}
        E_{i,i} & 0 \\
        0 & -E_{i,i}
    \end{pmatrix}
\end{equation}
where $E_{i,i}$ is a matrix with a nonzero value only in the $i$-th diagonal element:
\begin{equation}
    E_{i,i}=
    \begin{pmatrix}
        0&&&&&& \\
        &\ddots&&&&& \\
        &&0&&&& \\
        &&&1^{(i,i)}&&& \\
        &&&&0&& \\
        &&&&&\ddots& \\
        &&&&&&0
    \end{pmatrix}
\end{equation}
Thus, the maximal torus is given by:
\begin{equation}
    T=
    \begin{pmatrix}
        t & 0 \\
        0 & t^{-1}
    \end{pmatrix}
\end{equation}
where
\begin{equation}
    t=
    \begin{pmatrix}
        e^{i\theta_1}&&\\
        &\ddots&\\
        &&e^{i\theta_k}
    \end{pmatrix}
\end{equation}
The Weyl group of $Sp(k)$ is isomorphic to the Weyl group of $\mathfrak{sp}(k)$, so the action of the Weyl group on $Sp(k)$ can be understood by considering two types of elements:

\noindent
(1) The exchange of the $i$-th and $j$-th basis vectors, $U_{(i\leftrightarrow j)}\in Sp(k)$:
\begin{equation}
    U_{(i\leftrightarrow j)}=
    \begin{pmatrix}
        u_{(i\leftrightarrow j)}&0\\
        0&u_{(i\leftrightarrow j)}
    \end{pmatrix}
\end{equation}
where
\begin{equation}
    u_{(i\leftrightarrow j)}=
    \left(
    \begin{array}{ccccccccccc}
        1&&&&&&&&&&\\
        &\ddots&&&&&&&&&\\
        &&1&&&&&&&&\\
        &&&0&\cdots&\cdots&\cdots&1^{(i,j)}&&&\\
        &&&\vdots&1&&&\vdots&&&\\
        &&&\vdots&&\ddots&&\vdots&&&\\
        &&&\vdots&&&1&\vdots&&&\\
        &&&1^{(j,i)}&\cdots&\cdots&\cdots&0&&&\\
        &&&&&&&&1&&\\
        &&&&&&&&&\ddots&\\
        &&&&&&&&&&1
    \end{array}
    \right)
\end{equation}
The transformation induced by this matrix is
\begin{equation}
    U_{(i\leftrightarrow j)}TU_{(i\leftrightarrow j)}^\top\in T(\theta_i\leftrightarrow\theta_j)
\end{equation}
where the basis vectors of the maximal torus are swapped.

\noindent
(2) The reversal of the $i$-th basis vector, $U_{(i)}\in Sp(k)$:
\begin{equation}
    U_{(i)}=
    \left(
    \begin{array}{ccccccc|ccccccc}
        1&&&&&&&&&&&&&\\
        &\ddots&&&&&&&&&&&&\\
        &&1&&&&&&&&&&&\\
        &&&0&\cdots&\cdots&\cdots&\cdots&\cdots&\cdots&1^{(i,i+k)}&&&\\
        &&&\vdots&1&&&&&&\vdots&&&\\
        &&&\vdots&&\ddots&&&&&\vdots&&&\\
        &&&\vdots&&&1&&&&\vdots&&&\\
        \hline
        &&&\vdots&&&&1&&&\vdots&&&\\
        &&&\vdots&&&&&\ddots&&\vdots&&&\\
        &&&\vdots&&&&&&1&\vdots&&&\\
        &&&-1^{(i+k,i)}&\cdots&\cdots&\cdots&\cdots&\cdots&\cdots&0&&&\\
        &&&&&&&&&&&1&&\\
        &&&&&&&&&&&&\ddots&\\
        &&&&&&&&&&&&&1
    \end{array}
    \right)
\end{equation}
The transformation induced by this matrix is:
\begin{equation}
    U_{(i)}TU_{(i)}^\top\in T(\theta_i\to-\theta_i)
\end{equation}
which results in the inversion of the basis vector of the maximal torus. 

By considering the two types of transformations above, we can investigate the action of the Weyl group of the $Sp(k)$ group.
Let us examine the action of the Weyl group on the antisymmetric representation.
Here, we are interested in the action on the classical moduli space as shown in \ref{Eq:moduli_Sp_asym}, so we will only consider the eigenvalues $a_i$.
The eigenvalues are exchanged by $U_{(i\leftrightarrow j)}$:
\begin{equation}
    U_{(i\leftrightarrow j)}AU_{(i\leftrightarrow j)}^\top=A(a_i\leftrightarrow a_j)
\end{equation}
However, under $U_{(i)}$, the eigenvalues remain invariant:
\begin{equation}
    U_{(i)}AU_{(i)}^\top=A
\end{equation}
Thus, the action of the Weyl group on the classical moduli space of the antisymmetric representation is given by the permutation group $S_k$.

Following a similar procedure as above, we can determine the action of the Weyl group on the classical moduli space of the antisymmetric representation as shown in \ref{Eq:SU(2n)_moduli}.
The Weyl group of $SU(2k)$ is $S_{2k}$, so the action of the Weyl group is given by
\begin{subequations}
\begin{align}
    u_{(i\leftrightarrow j)}Au_{(i\leftrightarrow j)}^\top \\
    u_{(i\leftrightarrow j)}\tilde{A}u_{(i\leftrightarrow j)}^\top
\end{align}
\end{subequations}
where
\begin{equation}
    u_{(i\leftrightarrow j)}=
    \left(
    \begin{array}{ccccccccccc}
        1&&&&&&&&&&\\
        &\ddots&&&&&&&&&\\
        &&1&&&&&&&&\\
        &&&0&\cdots&\cdots&\cdots&1^{(i,j)}&&&\\
        &&&\vdots&1&&&\vdots&&&\\
        &&&\vdots&&\ddots&&\vdots&&&\\
        &&&\vdots&&&1&\vdots&&&\\
        &&&1^{(j,i)}&\cdots&\cdots&\cdots&0&&&\\
        &&&&&&&&1&&\\
        &&&&&&&&&\ddots&\\
        &&&&&&&&&&1
    \end{array}
    \right)
\end{equation}
When the superpotential is expressed in terms of the eigenvalues of the classical moduli space, it appears in the form $a_i\tilde{a}_i$.
Therefore, it is useful to investigate the action of the Weyl group on $A\tilde{A}$, which is given by
\begin{equation}
    u_{(i\leftrightarrow j)}A\tilde{A}u_{(i\leftrightarrow j)}^\top
\end{equation}
Since $A\tilde{A}$ is a diagonal matrix, the action of the Weyl group is induced by the permutation group $S_k$.
That is, the transformation of the eigenvalues of the classical moduli space, $a_i\tilde{a}_i\to a_j\tilde{a}_j$, corresponds to a gauge equivalent configuration.

\section{$s$-Confining Theories}
\label{sec:sconfining}

Here we show that none of the $s$-confining theories listed in \cite{Csaki:1996zb} is $t$-confining. 

We copied the table of candidate $s$-confining $SU$ theories in Table \ref{tab:SUtable}. It is clear that all $s$-confining theories have at least one field in the fundamental representation, and hence all center charges can be screened. The same is true with $s$-confining $Sp$ theories in Table \ref{tab:Sptable}.

\begin{table}[h]
\begin{center}
\renewcommand{\arraystretch}{1.8}
\begin{tabular}{|c|c|} \hline
$SU(N)$ & $(N+1) (\Yfund + \overline{\Yfund})$\\
$SU(N)$ & $\Yasymm + N\, \overline{\Yfund} + 4\, \Yfund $\\
$SU(N)$ & $\Yasymm + \overline{\Yasymm} + 3 (\Yfund + \overline{\Yfund})$ \\
\hline
$SU(5)$ & $ 3 (\Yasymm + \overline{\Yfund}) $ \\
$SU(5)$ & $ 2\, \Yasymm + 2\, \Yfund + 4\, \overline{\Yfund}$ \\
$SU(6)$ & $2\, \Yasymm + 5\, \overline{\Yfund} + \Yfund$ \\
$SU(6)$ & $\Ythreea + 4 (\Yfund + \overline{\Yfund})$ \\
$SU(7)$ & $2 (\Yasymm + 3\, \overline{\Yfund})$ \\
  \hline \end{tabular}
\end{center}
\caption{All $s$-confining $SU$ theories from \cite{Csaki:1996zb}.}
\label{tab:SUtable}
\end{table}

\begin{table}[h]
\begin{center}
\renewcommand{\arraystretch}{1.8}
\begin{tabular}{|c|c|} \hline
$Sp(2N)$ & $(2N+4)\, \Yfund$\\
$Sp(2N)$ & $\Yasymm +6\, \Yfund $\\
\hline
\end{tabular}
\end{center}
\caption{All $s$-confining $Sp$ theories from \cite{Csaki:1996zb}.}
\label{tab:Sptable}
\end{table}

The case of $SO$ theories requires more discussions. The list is given in Table \ref{tab:SOtable}. For $SO({\rm odd})$ groups, the center is $\bZ_2$, and the spinor representation is faithful under it. Therefore, the presence of a spinor representation makes it not $t$-confining. For $SO(4k+2)$ groups, the center is $\bZ_4$, and the spinor representation is again faithful under it. Therefore, the presence of a spinor representation makes it not $t$-confining. For $SO(4k)$ groups, however, the center is $\bZ_2 \times \bZ_2$, and $s$ can screen the first $\bZ_2$ but not the second, and $s'$ is the other way around. $v$ can screen the diagonal subgroup of two $\bZ_2$s. If at least two among the three presentations are present, therefore, all centers can be screened. Looking at the list of $SO(12)$ and $SO(8)$, that is the case. Namely none of them is $t$-confining.

\begin{table}[h]
\vspace*{-1cm}
\[
\renewcommand{\arraystretch}{1.2}
\begin{array}{|c|c|} \hline
SO(14) & (1,0,5)\\
SO(13) & (1,4)\\
SO(12) & (1,0,7)\\
SO(12) & (2,0,3)\\
SO(12) & (1,1,3)\\
SO(11) & (1,6)\\
SO(11) & (2,2)\\
SO(10) & (4,0,1)\\
SO(10) & (3,0,3)\\
SO(10) & (2,0,5)\\
SO(10) & (3,1,1)\\
SO(10) & (2,1,3)\\
SO(10) & (1,1,5)\\
SO(10) & (2,2,1)\\
SO(9) & (4,0)\\
SO(9) & (3,2)\\
SO(9) & (2,4)\\
SO(8) & (4,3,0)\\
SO(8) & (4,2,1)\\
SO(8) & (3,3,1)\\
SO(8) & (3,2,2)\\
SO(7) & (6,0)\\
SO(7) & (5,1)\\
SO(7) & (4,2)\\
SO(7) & (3,3)\\
\hline
\end{array}
\]
\caption{All $s$-confining $SO$ theories from \cite{Csaki:1996zb}. The field content is shown by the numbers in $(s,s',v)$ representations for $SO({\rm even})$ or $(s,v)$ representations for $SO({\rm odd})$.}
\label{tab:SOtable}
\end{table}

\section{Lie algebras}
\label{sec:data_of_Lie_algebra}

Here, we provide a summary of the Lie algebra data used in this paper.
For classical Lie algebras, we have compiled the Dynkin indices of each irreducible representation, and for $SU$-algebra, we have also included the anomalies.

\clearpage
\begin{table}[H]
    \ytableausetup{boxsize=2mm}
    \centering
    \myspace{1.3}
    \begin{tabular}{ScScSc}
        \hline
        representation & dimension & Dynkin index \\
        \hline
        $\adj$ & $n^2-1$ & $2n$ \\
        $\ydiagram{1}$ & $n$ & $1$ \\
        $\ydiagram{2}$ & $\dfrac{n(n+1)}{2}$ & $n+2$ \\
        $\ydiagram{1,1}$ & $\dfrac{n(n-1)}{2}$ & $n-2$ \\
        $\ydiagram{3}$ & $\dfrac{n(n+1)(n+2)}{6}$ & $\dfrac{(n+2)(n+3)}{2}$ \\
        $\ydiagram{2,1}$ & $\dfrac{n(n-1)(n+1)}{3}$ & $n^2-3$ \\
        $\ydiagram{1,1,1}$ & $\dfrac{n(n-1)(n-2)}{6}$ & $\dfrac{(n-3)(n-2)}{2}$ \\
        $\ydiagram{4}$ & $\dfrac{n(n+1)(n+2)(n+3)}{24}$ & $\dfrac{(n+2)(n+3)(n+4)}{6}$ \\
        $\ydiagram{3,1}$ & $\dfrac{n(n-1)(n+1)(n+2)}{8}$ & $\dfrac{(n+2)(n^2+n-4)}{2}$ \\
        $\ydiagram{2,2}$ & $\dfrac{n^2(n+1)(n-1)}{12}$ & $\dfrac{n(n-2)(n+2)}{3}$ \\
        $\ydiagram{2,1,1}$ & $\dfrac{n(n+1)(n-1)(n-2)}{8}$ & $\dfrac{(n-2)(n^2-n-4)}{2}$ \\
        $\ydiagram{1,1,1,1}$ & $\dfrac{n(n-1)(n-2)(n-3)}{24}$ & $\dfrac{(n-2)(n-3)(n-4)}{6}$ \\
        $\ydiagram{5}$ & $\dfrac{n(n+1)(n+2)(n+3)(n+4)}{120}$ & $\dfrac{(n+2)(n+3)(n+4)(n+5)}{24}$ \\
        $\ydiagram{4,1}$ & $\dfrac{n(n-1)(n+1)(n+2)(n+3)}{30}$ & $\dfrac{(n+2)(n+3)(n^2+2n-5)}{6}$ \\
        $\ydiagram{3,2}$ & $\dfrac{n^2(n-1)(n+1)(n+2)}{24}$ & $\dfrac{n(n+2)(5n^2+4n-25)}{24}$ \\
        $\ydiagram{3,1,1}$ & $\dfrac{n(n-2)(n-1)(n+1)(n+2)}{20}$ & $\dfrac{(n-2)(n+2)(n^2-5)}{4}$ \\
        $\ydiagram{2,2,1}$ & $\dfrac{n^2(n-2)(n-1)(n+1)}{24}$ & $\dfrac{n(n-2)(5n^2-4n-25)}{24}$ \\
        $\ydiagram{2,1,1,1}$ & $\dfrac{n(n-3)(n-2)(n-1)(n+1)}{30}$ & $\dfrac{(n-3)(n-2)(n^2-2n-5)}{6}$ \\
        $\ydiagram{1,1,1,1,1}$ & $\dfrac{n(n-4)(n-3)(n-2)(n-1)}{120}$ & $\dfrac{(n-5)(n-4)(n-3)(n-2)}{24}$ \\
        \hline
    \end{tabular}
    \caption{irreducible representaions of $SU(n)$}
\end{table}

\begin{table}[H]
    \ytableausetup{boxsize=2mm}
    \centering
    \myspace{1.3}
    \begin{tabular}{ScScSc}
        \hline
        representation & dimension & Dynkin index \\
        \hline
        $\ydiagram{1}$ & $2n$ & $1$ \\
        $\ydiagram{2}$ & $n(2n+1)$ & $2(n+1)$ \\
        $\ydiagram{1,1}$ & $n(2n-1)-1$ & $2(n-1)$ \\
        $\ydiagram{3}$ & $\dfrac{2n(n+1)(2n+1)}{3}$ & $(n+1)(2n+3)$ \\
        $\ydiagram{2,1}$ & $\dfrac{8n(n-1)(n+1)}{3}$ & $4(n+1)(n-1)$ \\
        $\ydiagram{1,1,1}$ & $\dfrac{2n(n-2)(2n+1)}{3}$ & $(2n-1)(n-2)$ \\
        $\ydiagram{4}$ & $\dfrac{n(n+1)(2n+1)(2n+3)}{6}$ & $\dfrac{2(n+1)(n+2)(2n+3)}{3}$ \\
        $\ydiagram{3,1}$ & $\dfrac{n(n-1)(2n+1)(2n+3)}{2}$ & $2(n-1)(n+1)(2n+3)$ \\
        $\ydiagram{2,2}$ & $\dfrac{n(n-1)(2n-1)(2n+3)}{3}$ & $\dfrac{2(n-1)(2n-1)(2n+3)}{3}$ \\
        $\ydiagram{2,1,1}$ & $\dfrac{(n-2)(n+1)(2n-1)(2n+1)}{2}$ & $2(n-2)(n+1)(2n-1)$ \\
        $\ydiagram{1,1,1,1}$ & $\dfrac{n(n-3)(2n-1)(2n+1)}{6}$ & $\dfrac{2(n-3)(n-1)(2n-1)}{3}$ \\
        \hline
    \end{tabular}
    \caption{irreducible representaions of $Sp(n)$}
\end{table}

\begin{table}[H]
    \ytableausetup{boxsize=2mm}
    \centering
    \myspace{1.3}
    \begin{tabular}{ScScSc}
        \hline
        representation & dimension & Dynkin index \\
        \hline
        $\mathrm{S}$ & $2^{\lfloor\frac{n-1}{2}\rfloor}$ & $2^{\lfloor\frac{n-5}{2}\rfloor}$ \\
        $\ydiagram{1}$ & $n$ & $2$ \\
        $\ydiagram{2}$ & $\dfrac{(n+2)(n-1)}{2}$ & $2(n+2)$ \\
        $\ydiagram{1,1}$ & $\dfrac{n(n-1)}{2}$ & $2(n-2)$ \\
        $\ydiagram{3}$ & $\dfrac{n(n-1)(n+4)}{6}$ & $(n+1)(n+4)$ \\
        $\ydiagram{2,1}$ & $\dfrac{n(n-2)(n+2)}{3}$ & $2(n-2)(n+2)$ \\
        $\ydiagram{1,1,1}$ & $\dfrac{n(n-1)(n-2)}{6}$ & $(n-2)(n-3)$ \\
        \hline
    \end{tabular}
    \caption{irreducible representaions of $SO(n)$}
\end{table}

\begin{table}[H]
    \ytableausetup{boxsize=2mm}
    \centering
    \myspace{1.4}
    \begin{tabular}{ScSc}
        \hline
        representaion & anomaly \\
        \hline
        $\adj$ & $0$ \\
        $\ydiagram{1}$ & $1$ \\
        $\ydiagram{2}$ & $n+4$ \\
        $\ydiagram{1,1}$ & $n-4$ \\
        $\ydiagram{3}$ & $\dfrac{(n+3)(n+6)}{2}$ \\
        $\ydiagram{2,1}$ & $n^2-9$ \\
        $\ydiagram{1,1,1}$ & $\dfrac{(n-3)(n-6)}{2}$ \\
        $\ydiagram{4}$ & $\dfrac{(n+3)(n+4)(n+8)}{6}$ \\
        $\ydiagram{2,2}$ & $\dfrac{n(n-4)(n+4)}{3}$ \\
        $\ydiagram{2,1,1}$ & $\dfrac{(n-4)(n^2-n-8)}{2}$ \\ 
        \hline
    \end{tabular}
    \caption{anomalies of irreducible representaions of $SU(n)$}
\end{table}

\acknowledgments

RI and SS are supported by Forefront Physics and Mathematics Program to Drive Transformation (FoPM), a World-leading Innovative Graduate Study (WINGS) Program, the University of Tokyo. SS is also supported by Research Fellow of Japan Society for the Promotion of Science (JSPS Research Fellow), JSPS KAKENHI Grant Number JP25KJ0857, and JSR Fellowship, the University of Tokyo. The work of HM is supported by the Director, Office of Science, Office of High Energy Physics of the U.S. Department of Energy under the Contract No. DE-AC02-05CH11231, by the NSF grant PHY-2210390, by the JSPS Grant-in-Aid for Scientific Research JP23K03382, MEXT Grant-in-Aid for Transformative Research Areas (A) JP20H05850, JP20A203, Hamamatsu Photonics, K.K, and Tokyo Dome Corportation. In addition, HM is supported by the World Premier International Research Center Initiative (WPI) MEXT, Japan.


\bibliography{./bibliography}
\bibliographystyle{JHEP}

\end{document}